   \documentstyle[12pt]{article}
\if@twoside\oddsidemargin25mm
\else\oddsidemargin30mm\evensidemargin30mm\marginparwidth30mm\fi%
\begin{document}
\font\cmss=cmss10 \font\cmsss=cmss10 at 7pt
\def\twomat#1#2#3#4{\left(\matrix{#1 & #2 \cr #3 & #4}\right)}
\def\inbar{\vrule height1.5ex width.4pt depth0pt}
\def\IC{\relax\,\hbox{$\inbar\kern-.3em{\rm C}$}}
\def\ID{\relax{\rm I\kern-.18em D}}
\def\IF{\relax{\rm I\kern-.18em F}}
\def\IH{\relax{\rm I\kern-.18em H}}
\def\II{\relax{\rm I\kern-.17em I}}
\def\IN{\relax{\rm I\kern-.18em N}}
\def\IP{\relax{\rm I\kern-.18em P}}
\def\IQ{\relax\,\hbox{$\inbar\kern-.3em{\rm Q}$}}
\def\bfzero{\relax\,\hbox{$\inbar\kern-.3em{\rm 0}$}}
\def\IR{\relax{\rm I\kern-.18em R}}
\def\ZZ{\relax\ifmmode\mathchoice
{\hbox{\cmss Z\kern-.4em Z}}{\hbox{\cmss Z\kern-.4em Z}}
{\lower.9pt\hbox{\cmsss Z\kern-.4em Z}}
{\lower1.2pt\hbox{\cmsss Z\kern-.4em Z}}\else{\cmss Z\kern-.4em
Z}\fi}
\def\bfone{\relax{\rm 1\kern-.35em 1}}
\def\dop{{\rm d}\hskip -1pt}
\def\real{{\rm Re}\hskip 1pt}
\def\ee#1{{\rm e}^{#1}}
\def\trace{{\rm Tr}\hskip 1pt}
\def\ii{{\rm i}}
\def\diag{{\rm diag}}
\def\sch#1#2{\{#1;#2\}}
\newcommand{\eqn}[1]{(\ref{#1})}
\newcommand{\ft}[2]{{\textstyle\frac{#1}{#2}}}
\newcommand{\QED}{{\hspace*{\fill}\rule{2mm}{2mm}\linebreak}}
\renewcommand{\theequation}{\thesection.\arabic{equation}}
\csname @addtoreset\endcsname{equation}{section}
\newsavebox{\uuunit}
\sbox{\uuunit}
{\setlength{\unitlength}{0.825em}
 \begin{picture}(0.6,0.7)
\thinlines
\put(0,0){\line(1,0){0.5}}
\put(0.15,0){\line(0,1){0.7}}
 \put(0.35,0){\line(0,1){0.8}}
\multiput(0.3,0.8)(-0.04,-0.02){12}{\rule{0.5pt}{0.5pt}}
\end {picture}}
\newcommand {\unity}{\mathord{\!\usebox{\uuunit}}}
\newcommand  {\Rbar} {{\mbox{\rm$\mbox{I}\!\mbox{R}$}}}
\newcommand  {\Hbar} {{\mbox{\rm$\mbox{I}\!\mbox{H}$}}}
\newcommand {\Cbar}
{\mathord{\setlength{\unitlength}{1em}
 \begin{picture}(0.6,0.7)(-0.1,0)
  \put(-0.1,0){\rm C}
   \thicklines
  \put(0.2,0.05){\line(0,1){0.55}}
    \end {picture}}}
\newsavebox{\zzzbar}
\sbox{\zzzbar}
  {\setlength{\unitlength}{0.9em}
  \begin{picture}(0.6,0.7)
 \thinlines
  \put(0,0){\line(1,0){0.6}}
  \put(0,0.75){\line(1,0){0.575}}
\multiput(0,0)(0.0125,0.025){30}{\rule{0.3pt}{0.3pt}}
  \multiput(0.2,0)(0.0125,0.025){30}{\rule{0.3pt}{0.3pt}}
\put(0,0.75){\line(0,-1){0.15}}
  \put(0.015,0.75){\line(0,-1){0.1}}
  \put(0.03,0.75){\line(0,-1){0.075}}
 \put(0.045,0.75){\line(0,-1){0.05}}
 \put(0.05,0.75){\line(0,-1){0.025}}
 \put(0.6,0){\line(0,1){0.15}}
  \put(0.585,0){\line(0,1){0.1}}
  \put(0.57,0){\line(0,1){0.075}}
 \put(0.555,0){\line(0,1){0.05}}
 \put(0.55,0){\line(0,1){0.025}}
  \end{picture}}
\newcommand{\Zbar}{\mathord{\!{\usebox{\zzzbar}}}}
\newcommand{\Ka}{K\"ahler}
\newcommand{\qu}{quaternionic}
\def\ib{{\bar \imath}}
\def\jb{{\bar \jmath}}
\def\bfone{\relax{\rm 1\kern-.35em 1}}
\font\cmss=cmss10 \font\cmsss=cmss10 at 7pt
\def\a{\alpha} \def\b{\beta} \def\d{\delta}
\def\e{\epsilon} \def\c{\gamma}
\def\G{\Gamma} \def\l{\lambda}
\def\L{\Lambda} \def\s{\sigma}
\def\cA{{\cal A}} \def\cB{{\cal B}}
\def\cC{{\cal C}} \def\cD{{\cal D}}
\def\cF{{\cal F}} \def\cG{{\cal G}}
\def\cH{{\cal H}} \def\cI{{\cal I}}
\def\cJ{{\cal J}} \def\cK{{\cal K}}
\def\cL{{\cal L}} \def\cM{{\cal M}}
\def\cN{{\cal N}} \def\cO{{\cal O}}
\def\cP{{\cal P}} \def\cQ{{\cal Q}}
\def\cR{{\cal R}} \def\cV{{\cal V}}\def\cW{{\cal W}}
%
%
%
\def\crr{\crcr\noalign{\vskip {8.3333pt}}}
\def\tilde{\widetilde}
\def\bar{\overline}
\def\us#1{\underline{#1}}
\let\shat=\hat
\def\hat{\widehat}
\def\hyp{\vrule height 2.3pt width 2.5pt depth -1.5pt}
\def\square{\mbox{.08}{.08}}
\def\Coeff#1#2{{#1\over #2}}
\def\Coe#1.#2.{{#1\over #2}}
\def\coeff#1#2{\relax{\textstyle {#1 \over #2}}\displaystyle}
\def\coe#1.#2.{\relax{\textstyle {#1 \over #2}}\displaystyle}
\def\half{{1 \over 2}}
\def\shalf{\relax{\textstyle {1 \over 2}}\displaystyle}
\def\dag#1{#1\!\!\!/\,\,\,}
\def\to{\rightarrow}
\def\notin{\hbox{{$\in$}\kern-.51em\hbox{/}}}
\def\shdot{\!\cdot\!}
\def\ket#1{\,\big|\,#1\,\big>\,}
\def\bra#1{\,\big<\,#1\,\big|\,}
\def\equaltop#1{\mathrel{\mathop=^{#1}}}
\def\Trbel#1{\mathop{{\rm Tr}}_{#1}}
\def\inserteq#1{\noalign{\vskip-.2truecm\hbox{#1\hfil}
\vskip-.2cm}}
\def\attac#1{\Bigl\vert
{\phantom{X}\atop{{\rm\scriptstyle #1}}\phantom{X}}}
\def\exx#1{e^{{\displaystyle #1}}}
\def\del{\partial}
\def\delbar{\bar\partial}
\def\nex#1{$N\!=\!#1$}
\def\dex#1{$d\!=\!#1$}
\def\cex#1{$c\!=\!#1$}
\def\eg{{\it e.g.}} \def\ie{{\it i.e.}}
\def\IE{\relax{{\rm I\kern-.18em E}}}
\def\cE{{\cal E}}
\def\rt{{\cR^{(3)}}}
\def\IGam{\relax{{\rm I}\kern-.18em \Gamma}}
\def\IGa{\IA}
\def\ii{{\rm i}}
\def\diag{{\rm diag}}
\def\hp{{\bf HP}^{4(m+3)}}
\def\omep{\omega^{\scriptscriptstyle +}}
\def\omepind#1{\omega^{\scriptscriptstyle +\hskip 2pt #1}}
\def\omem{\omega^{\scriptscriptstyle -}}
\def\omemind#1{\omega^{\scriptscriptstyle -\hskip 2pt #1}}
\def\omepm{\omega^{\scriptscriptstyle \pm}}
\def\omepmind#1{\omega^{\scriptscriptstyle \pm\hskip 2pt #1}}
\def\Omep{\Omega^{\scriptscriptstyle +}}
\def\Omepind#1{\Omega^{\scriptscriptstyle +\hskip 2pt #1}}
\def\Omem{\Omega^{\scriptscriptstyle -}}
\def\Omemind#1{\Omega^{\scriptscriptstyle -\hskip 2pt #1}}
\def\Omepm{\Omega^{\scriptscriptstyle \pm}}
\def\Omepmind#1{\Omega^{\scriptscriptstyle \pm\hskip 2pt #1}}
\def\Jp{{\cal J}^{\scriptscriptstyle +}}
\def\Jpind#1{{\cal J}^{\scriptscriptstyle +\hskip 2pt #1}}
\def\Jm{{\cal J}^{\scriptscriptstyle -}}
\def\Jmind#1{{\cal J}^{\scriptscriptstyle -\hskip 2pt #1}}
\def\Jpm{{\cal J}^{\scriptscriptstyle \pm}}
\def\Jpmind#1{{\cal J}^{\scriptscriptstyle \pm\hskip 2pt #1}}
\def\Jmp{{\cal J}^{\scriptscriptstyle \mp}}
\def\Jmpind#1{{\cal J}^{\scriptscriptstyle \mp\hskip 2pt #1}}
\def\Qpind#1{Q^{\scriptscriptstyle +\hskip 2pt #1}}
\def\Qmind#1{Q^{\scriptscriptstyle -\hskip 2pt #1}}
\def\Qpmind#1{Q^{\scriptscriptstyle -\hskip 2pt #1}}
\def\FFpind{F^{\scriptscriptstyle + }}
\def\FFmind{F^{\scriptscriptstyle -}}
\def\KKpind{K^{\scriptscriptstyle + }}
\def\KKmind{K^{\scriptscriptstyle -}}
\def\KKpmind{K^{\scriptscriptstyle \pm}}
\newtheorem{definizione}{Definition}
\newtheorem{domanda}{Question}
\newtheorem{risposta}{Answer}
\def\o#1#2{{#1}\over{#2}}
\newtheorem{affermazione}{Statement}
\begin{titlepage}
\hskip 5.5cm
\vbox{\hbox{SISSA 64/95/EP}\hbox{POLFIS-TH 07/95}
\hbox{CERN-TH/95-140}\hbox{UCLA/95/TEP/19}
}
\hskip 1.5cm
\vbox{\hbox{IFUM 508FT}\hbox{KUL-TF-95/18}
\hbox{hep-th/9506075}\hbox{June, 1995}}
\vfill
\begin{center}
{\LARGE A Search for Non-Perturbative Dualities \\
\vskip 1.5mm
 of Local $N=2$ Yang--Mills Theories\\
\vskip 3.5mm
  from Calabi--Yau Threefolds $^*$ }\\
\vfill
{\large M. Bill\'o$^1$, A. Ceresole$^2$,
R. D'Auria$^2$, S. Ferrara$^3$,\\
 \vskip 1.5mm
 P. Fr\'e$^1$,  T. Regge$^2$, P. Soriani$^4$
and  A. Van Proeyen$^{5}$ $^\dagger$ } \\
\vfill
{\small
$^1$ International School for Advanced Studies (ISAS), via Beirut 2-4,
I-34100 Trieste\\
and Istituto Nazionale di Fisica Nucleare (INFN) - Sezione di Trieste, Italy\\
\vspace{6pt}
$^2$ Dipartimento di Fisica, Politecnico di Torino,\\
 Corso Duca degli Abruzzi 24, I-10129 Torino\\
and Istituto Nazionale di Fisica Nucleare (INFN) - Sezione di Torino, Italy\\
\vspace{6pt}
$^3$ CERN Theoretical Division, CH 1211 Geneva, Switzerland\\
and UCLA Physics Department, Los Angeles CA, USA\\
\vspace{6pt}
$^4$ Dipartimento di Fisica, Universit\`a di Milano, via Celoria 6,
I-20133 Milano,\\
and Istituto Nazionale di Fisica Nucleare (INFN) - Sezione di Milano, Italy\\
\vspace{6pt}
$^5$ Instituut voor Theoretische Fysica - Katholieke Universiteit Leuven
\\Celestijnenlaan 200D B--3001 Leuven, Belgium\\
}
\end{center}
\vfill
\begin{center}
{\bf Abstract}
\end{center}
{\small
The generalisation of the rigid special geometry of the vector
multiplet quantum moduli space to the case of supergravity is
discussed through the notion of a dynamical Calabi--Yau threefold.
Duality symmetries of this manifold are connected with the
analogous dualities associated with the dynamical Riemann surface
of the rigid theory. N=2 rigid gauge theories are reviewed in a
framework ready for comparison with the local case. As a byproduct
we give in general the full duality group (quantum monodromy) for an
arbitrary rigid $SU(r+1)$ gauge theory, extending previous explicit
constructions for the $r=1,2$ cases. In the coupling to gravity,
R--symmetry and monodromy groups of the dynamical Riemann surface,
whose structure we discuss in detail, are embedded into the
symplectic duality group $\Gamma_D$ associated with the moduli space
of the dynamical Calabi--Yau threefold.}
\vspace{2mm} \vfill \hrule width 3.cm
{\footnotesize
\noindent $^\dagger$ Onderzoeksleider, NFWO, Belgium\\
\noindent
$^*$ Supported in part by DOE grant
DE-FGO3-91ER40662, Task C.
and by EEC Science Program SC1*CT92-0789.}
\end{titlepage}
\section{Introduction}
Recent progress \cite{SW1,SW2} towards understanding non-perturbative
properties of N=2 Yang--Mills theories has been obtained by associating the
holomorphic N=2 prepotential \cite{BDW} to the periods
of an auxiliary Riemann surface (of genus $r$ equal to the
rank of the gauge group $G=SU(r+1)$), where the monodromy group
is directly related to the electric--magnetic duality symmetries of the
theory\cite{kltold,faraggi}.
\par
Non-perturbative monodromies related to monopole point singularities
(i.e. points where particles with non-vanishing magnetic charges,
monopoles or dyons, become massless) correspond to an infinite sum
of instanton contributions to the prepotential in the microscopic
G-invariant theory.
\par
Perturbative monodromies, on the other hand, correspond to the unique
one-loop perturbative correction \cite{PDV}
to the prepotential in the original
G-invariant theory, broken down to $U(1)^r$.
\par
Very recently, these exact results for the low-energy effective
N=2 Yang--Mills theory have been extended to include gravity by using
several different informations \cite{CDF}-\cite{kava}.
Firstly, in a paper by some of us \cite{CDF}, it was pointed
out that, in the case of coupling Yang--Mills theories to gravity,
the N=2 rigid special geometry encompassing the moduli space of
hyperelliptic Riemann surfaces is drastically modified by the
gravitational effects, in particular by the
presence of the graviphoton.
The immediate consequence of this is a  change
of the electric-magnetic duality group and also of the argument of
positivity of the metric of the moduli space of locally supersymmetric
Yang--Mills theories. Indeed, the same argument
used in the rigid case to introduce an auxiliary Riemann surface
to solve the theory, strongly suggests that the auxiliary surface
should be, in this case, a Calabi--Yau threefold with third
Betti number $b_3 = 2n$ where $n$ is the
total number of vectors in the theory.\par
Thus it would follow immediately that the electric-magnetic duality
group $\Gamma_D$ for the gravitational case is related to
the monodromy group $\Gamma_M\subset Sp(b_3,\ZZ)$
of the Calabi--Yau threefold . However we
observe here a conceptual difference from the rigid theory,
that for $r=1$ was recently pointed out in \cite{lopez},
 namely that in the local
case the electric-magnetic duality is larger than in the rigid theory
because also the symmetries
of the Calabi--Yau threefold that are not in the monodromy group
will be
in $\Gamma_D$. The symmetries of the Calabi--Yau defining equation
are $\Gamma_D/\Gamma_M$. This has no analogue in the rigid theories,
except for those symmetries of the auxiliary Riemann surface
defining polynomial that correspond to a unimodular rescaling factor,
such as the R-symmetry \cite{noi}.
\par
In this paper we will study classes of Calabi--Yau threefolds which
are potential candidates to satisfy the important requirement  to
realize the embedding of the  auxiliary Riemann surfaces of the rigid
Yang--Mills theories.
This entails a suitable embedding of duality symmetries
and monodromies of the rigid case.
We will actually derive an explicit representation of the monodromy group
of the $SU(n)$ rigid theories by extending some techniques introduced in
ref. \cite{CDR}
and later used to study the monodromies of Calabi--Yau manifolds for more
than one modulus. (The  monodromy for the $SU(n)$ case has been recently
 analysed in \cite{kltnew}, where explicit results were given for the
$SU(3)$ theory).
\par
 The criterion of searching for the right embedding can be
naturally implemented, using a series of different recent results,
in the context of string theories.
There, it is natural to associate to a given model
some dual theory, where the dynamical Calabi--Yau manifold is
not just an auxiliary geometrical tool
suitable for the analysis of the quantum behaviour, but the
target space of
the dual theory.
This means that, at least in the (abelian) phase where the
electric-magnetic
duality is manifest, it should be possible to solve the original
theory (which is known only in the
region of weak coupling) in the strong coupling regime by means of
another theory in its semiclassical regime \cite{huto1,Wdy}.
It is natural to associate gravitationally coupled N=2 Yang--Mills
theories to heterotic strings having N=2 supersymmetry in D=4
\cite{CDF}, and to identify their dual theories with type-IIA (or B)
 theories
\cite{kltold,CDFVP} compactified on the appropriate Calabi--Yau
threefolds \cite{CDFVP,CDF}. In this (in principle) more restricted
framework the auxiliary Calabi--Yau threefold previously considered
should then be identified with the compactification manifold of the
dual type II theory. Thus we
have, as proposed in \cite{FHSV}, a second-quantized mirror
symmetry in the sense that,
for vector multiplets,  the {\it classical} moduli space of the
Calabi--Yau manifold in the dual theory should give the {\it quantum}
moduli space of the heterotic theory on
$K3\times T_2$. This is made possible by the peculiar role of the
dilaton-axion complex field $S$ in string theory.
On one side it plays the role of string
``coupling constant'', on the other side it sits in a vector multiplet in
heterotic theories and in a hypermultiplet in type-II theories
\cite{CFG}.
This has a two-fold consequence: using N=2 supersymmetry,
which forbids \cite{BDW,spec1,spec2} the mixing of neutral moduli
in vector multiplets
with those of hypermultiplets in the low-energy lagrangian, it allows
to extend to N=2 string theories powerful
non-renormalization theorems of renormalizable N=2 gauge theories
\cite{nonre}. In particular, on the heterotic side, the
classical hypermultiplet quaternionic manifold does not receive
any quantum correction
\cite{AFGNT,DWKLL}.
For the dual type-II theory, the same is true for the manifold
 of the vector
multiplets \cite{stromco}.
\par
Since the exact moduli space of vector multiplets can be
obtained by first-quantized mirror symmetry\cite{mirror},
then it follows that the full heterotic string moduli space of
vector multiplets, i.e. the perturbative + instanton
corrected prepotential is
given by a ``classical'' computation on the type-II side.
\par
This precisely realizes the fact that the Calabi--Yau threefold
moduli space generalizes, in the case of strings, the
auxiliary Riemann surface \cite{SW1,SW2}.
In this context we notice that the duality group of the
Calabi--Yau space, which corresponds to the quantum monodromy of
the heterotic strings, also realizes at the
N=2 level, the $U$-duality idea of Hull and Townsend
\cite{huto1} and the $S$-$T$ duality
of Duff \cite{Duff}, since $S_D$, the dual of $S$, must be one of
the Calabi--Yau moduli. Of course the BPS saturated states of
type-II theories must be non-perturbative since they must contain
electric and magnetically charged states
with respect to the $U(1)$ gauge field of the three-form cohomology.
These states can be interpreted as particular topological states
coming from the compactification of a three-brane soliton.
In the N=4 case where, unlike the N=2 case,
quantum corrections are absent, a  pair of naturally dual
theories is known, namely heterotic on $T_6$ and type-II
on $K3\times T_2$.
The spectrum of BPS states in N=4 heterotic (using string--three-brane
duals) has been studied by Sen and Schwarz \cite{SS} and the enhanced
symmetries of $K3\times T_2$ by Hull and Townsend \cite{huto1},
Witten\cite{Wdy}, Harvey and Strominger\cite{HS}.
\par
In string theories this duality pairing is also made possible by
two other important facts:
\begin{enumerate}
\item[{\it i)}] the realization that on Calabi--Yau manifolds one
can have a change of Hodge numbers by \cite{stromco,GMS} conifold
transitions, i.e. black hole condensation through VEV's of
hypermultiplets carrying Ramond-Ramond charges which lower the
rank of the gauge group and increase the number of neutral
hypermultiplets. This phenomenon is dual to enhanced symmetries
or monopole singularities in heterotic
theories which may also change the number of massless vector
 multiplets and hypermultiplets. This also allows to connect a web of
heterotic theories to a web of Calabi--Yau
manifolds, through non-perturbative black-hole
condensation or monopole point singularities.
\item[{\it ii)}] Some Calabi--Yau manifolds can also be obtained
\cite{FHSV} by a
$\ZZ_2$ modding of the $K3\times T_2$
manifold which is in turn the type-II manifold yielding the
 model dual to the heterotic string compactified on $T_6$
in six dimensions\cite{HS}.
This is also the approach which allows to make an explicit
construction of the dual Calabi--Yau
manifold and then can be possibly extended to more general cases
by conifold transitions.
\end{enumerate}
Coming back to our original motivation to use a Calabi--Yau
threefold to embed the dynamical Riemann surface of rigid theories,
an important restriction comes from
the fact that the intersection form $d_{ABC}$ of this Calabi--Yau
should have the properties:
\begin{equation}
d_{ABC}=\{d_{Sij}\ ,\ d_{ijk}\}\ ,\ \ d_{ABC}=0\ {\rm otherwise},\ \
 A=\{S,i\}\ , \ i=1\ldots ,r
\end{equation}
where $S$ is the four dimensional dilaton-axion field and $d_{Sij}=C_{ij}$
corresponds to the tree level term of the  prepotential of
heterotic string vacua on $K_3\times
T_2$,
\begin{equation}
\label{int1}
{\cal F} (S,t_i) = S C_{ij} t^i t^j,
\hskip 1.5cm d_{Sij}= C_{ij} \not= 0\ .
\end{equation}
 $C_{ij} t^i t^j$ is a quadratic real form with signature $(1,r-1)$
and $d_{Sij}$ is the intersection form of the special
homogeneous spaces $SU(1,1)/U(1)\times
O(2,r)/O(2)\times O(r)$.
Non-vanishing $d_{ijk}$ terms can be induced by one loop contributions
on the heterotic side and
correspond to non--renormalizable terms suppressed by an inverse power
of $M_P^2$ in the rigid theory.
The previous requirement on $d_{ABC}$ poses a strong constraint on the
Calabi--Yau manifold, and a list
of such manifolds will be given for $r\leq 2$\footnote{ Calabi--Yau
manifolds obtained as $K3$ fibrations have been shown to satisfy
automatically the above constraint\cite{KLM}.}
\par
Recently Kachru and Vafa \cite{kava} enumerated Calabi--Yau threefolds
which are potential duals of heterotic strings with a given number of
vector and hypermultiplets
(in their abelian phase). Only manifolds for which the two distinct
types of multiplets agree  with known examples of heterotic string
 are considered. Moreover in some cases these authors give indications
that also terms in the prepotential with $S$ large but $t_i$ finite (such as
$\lim_{S\to\infty} \del_{ijk} {\cal F}(t,S)$)
agree with the pole structure
expected from one loop calculations in heterotic theory
\cite{AFGNT,DWKLL}.
\par
Our analysis is somewhat complementary,
in the sense that it does not focus
on the number of multiplets but rather on the possibility that a given
Calabi--Yau manifold can embed the monodromies of the Riemann surface of
the rigid theory, in presence of a dynamical coupling constant (dilaton).
\par
This last fact gives us the condition on the intersection
numbers which should possibly restrict the search for ``dual''
Calabi--Yau threefolds, especially
with $r$ large.
It is in fact rather obvious that the larger is $r$,
the more Calabi--Yau manifolds exist, but the
more stringent will probably look the constraints on the
intersection matrices. Thus,
the potential Calabi--Yau candidates will perhaps
not be too many (if any).
\par
The other important embedding is the R--symmetry, related to
non--perturbative corrections. We identify such symmetry with the
$\ZZ_p$ symmetry which usually occurs in Calabi--Yau threefolds. Once
this identification is made, one can explore how it acts on the
vector multiplets of the theory. This will be exhibited in a
particular example in section~\ref{ss:SGCY}. The conjecture that
suitable Calabi--Yau manifolds are non--perturbative solutions of
$N=2$ local Yang--Mills theories (or heterotic $N=2$ strings), should
pass the stringent test that such R--symmetries should find an
explanation in terms of the instanton configurations which occur in
gravitationally coupled Yang--Mills theories. Relations between
space--time instantons and world--sheet instanton sums have been
recently discussed for a two parameter Calabi--Yau threefold in
\cite{GomLopII}. We will just observe that in some cases such
R--symmetry can be explained by noticing that the Calabi--Yau
manifold contains two dimensional submanifolds realizing a multiple
cover of a Riemann surface identical in form with that occurring in
the rigid theory.
\par
It is important to stress that our search for dynamical
Calabi--Yau manifolds, without imposing a string duality,
is neither more nor less general than the
counting of Hodge numbers made in \cite{kava} by
imposing string duality only.
The reason is that the identification of Calabi--Yau dual to
heterotic strings
by just matching the number of vectors and hypermultiplets, without
imposing a priori the constraints from the quantum monodromy and the
intersection matrices, seems a necessary but not sufficient
assumption.
On the other hand, our criterion of
searching for quantum Calabi--Yau manifolds does not guarantee a
priori that the number of neutral hypermultiplets, in the abelian phase
of the heterotic theory, is the correct one.
The analysis seem to almost completely overlap for $r=1,2$ but we expect
them to be different and complementary for large $r$.
In particular, in \cite{kava} it is found the heterotic
realization of four out of the five\footnote{The published version
of \cite{kava} does not mention all these realizations.}
Calabi--Yau manifolds  with $h^{1,1}=2$ ($r=1$)
and of two of the manifolds with
$h^{1,1}=3$ that we have listed in
eq. (\ref{tabella}).
In Section 3, we analyze in particular two of the models with $r=1$
having heterotic counterparts, for which, thanks to the results of
\cite{cand}, the test can be done completely (writing explicitly also
the $Sp(6,\ZZ)$ duality matrices) in one case, and almost completely in the
other.
\par
Interestingly enough, in ref. \cite{kava} also models with $h^{1,1}=1$
 (or $h^{2,1}=1$ in the Calabi--Yau mirror) were analyzed. These
Type II theories would correspond, on the heterotic side, to theories with
only the graviphoton and dilaton vector giving a pure gravitational
$U(1)^2$. In this case, the BPS states would correspond to gravitational
states such as black holes and H-monopoles\footnote
{Curiously in these theories,
in the weak coupling regime, an unconventional three-dilaton vertex is
predicted, whose meaning deserves attention.}.
\par
Even more intriguing are Calabi--Yau manifolds in Type IIB theories with
$h^{2,1}=0$, whose mirror is not a Calabi--Yau\cite{h21zero}.
 This gives a prepotential ${\cal F}(X^0)=S_0
(X^0)^2$ , where $S_0$ is a complex constant.
These models would correspond, by duality, to a frozen dilaton
$S\equiv S_0=\frac{i}{g^2}+\theta$. In this case the BPS states may have
exact $Sp(2,\ZZ)$ duality, similar to $N=4$ rigid Yang--Mills theory,
because the graviphoton would just have a field independent complex
constant ${\rm Im} S_0$ and $\theta$-angle, ${\rm Re} S_0$.
\par
It would be extremely interesting to see whether
requiring both the quantum
consistency and the matching of the number of
supermultiplets one could obtain a unique and perhaps universal
classification.
\vskip 0.1cm
This paper is organized as follows. Section 2 is devoted to
rigid theories. We show how to define
the auxiliary Riemann surfaces that encode the exact solution of the
rigid N=2 gauge theories as hypersurfaces in weighted projective spaces.
Symmetry of the potential, monodromy, duality and the emerging of
the $\ZZ_{2r+2}$ R-symmetry group
are discussed along this line. In section 3 the structure of the
monodromy group for rigid $SU(r+1)$ theories is derived as a
subgroup of the braid group $B(2r+2)$.
In section 4 we consider local N=2 theories. The notion
of a dynamical Calabi--Yau manifold and the requirements that
such a manifold should satisfy in order to embed correctly the rigid
theories are discussed, and examples are given. In particular we stress
the role of the intersection numbers (corresponding to the structure constants
of the chiral ring) and the embedding of the R-symmetry group.
 In section 5 the definition of
central charge in type-II strings on Calabi--Yau threefolds is
discussed;
this is a basic point in showing  how the
Seiberg-Witten construction is realized at the string level
under the mapping to the dual heterotic theories. Finally,
the Appendix  contains additional remarks about
the rigid $SU(2)$ theory, mainly concerning the actual role of the
symmetries of the potential.
\section{Non-perturbative solutions of
Rigid N=2 gauge theories revisited}
Let us summarize the results obtained for pure N=2
gauge theories \cite{SW1,kltold,faraggi,kltnew},
 without hypermultiplets coupling.
For the N=2 {\it microscopic
gauge theory} of a group $G$, with Lie algebra ${\bf G}$
the rigid special geometry is encoded in a
``minimal coupling" quadratic prepotential
of the form:
\begin{eqnarray}
{\cal F}^{(micro)}(Y)&=&g_{IJ}^{(K)} \, Y^I \,
Y^J\nonumber\\
g_{IJ}^{(K)}&=&\mbox{Killing metric of
the Lie algebra}~{\bf G}
\label{mincoup_1}
\end{eqnarray}
This choice is motivated by renormalizability, positivity of
the energy and canonical normalization of the kinetic terms.
Consider next the effective lagrangian describing the dynamics
of the massless modes. This  is an abelian
 N=2 gauge theory that admits the maximal torus
${H}\subset{G}$
as new gauge group and is based on a new rigid special geometry:
\begin{eqnarray}
{\cal F}^{(eff)}(Y)&=&g_{\alpha\beta}^{(K)} \, Y^\alpha \,
Y^\beta + \Delta{\cal F}^{(eff)}\left ( Y^\alpha \right)\nonumber\\
Y^\alpha&\in& {\bf H} \subset {\bf G}
\end{eqnarray}
In general the effective prepotential ${\cal F}^{(eff)}(Y)$ has
a transcendental dependence on the scalar fields $Y^\alpha$
of the Cartan subalgebra multiplets, due to the
correction $\Delta{\cal F}^{(eff)}\left ( Y^\alpha \right)$.
The main problem is the determination of this correction.
Perturbatively one can get information on
$\Delta{\cal F}^{(eff)}\left ( Y^\alpha \right)$ and discover its
logarithmic singularity for large values of the scalar fields
$Y^\alpha$.
In particular one has a logarithmic correction to the gauge coupling
matrix
\begin{equation}
\Delta {\bar {\cal N}}_{\alpha \beta}=
{{\partial}^2 \over{ \partial Y^\alpha \partial Y^\beta}}
\Delta {\cal F}\ \stackrel{Y \to \infty}\sim \ \sum_{\bf\alpha}
{\bf\alpha}^\alpha {\bf\alpha}^\beta
\log{(Y \cdot {\bf\alpha})^2\over \Lambda^2}
\end{equation}
where ${\bf\alpha}$ are the root vectors of the gauge Lie algebra and
$\Lambda^2$ is the dynamically generated scale.
The perturbative monodromy following from
\begin{eqnarray}
{\cal N}_{\alpha \beta} &\rightarrow &
[(C+ D{\cal N})(A + B{\cal N})^{-1}]_{\alpha\beta} \nonumber \\
Sp(2r, \IR) &\ni & \twomat{A}{B}{C}{D} \sim
\twomat{1}{0}{ \sum_{\bf\alpha} {\bf\alpha}^\alpha {\bf\alpha}^\beta}
{1}
\end{eqnarray}
is assumed to be a part of the monodromy group
of a genus $r$ Riemann surface having a symplectic action on
 the periods
of the surface. Guessing such a dynamical Riemann surface gives
the nonperturbative structure $\Delta {\cal F}^{(eff.)}$.
\par
Denoting by $r$ the rank of the original
gauge group ${G}$, one derives the structure of the
effective gauge theory of the maximal torus ${\bf H}$ from
the geometry of an $r$--parameter family ${\cal M}_{1}[r]$
of dynamical genus $r$ Riemann surfaces. The essential steps
of the procedure are as follows: naming $u_i$
(i=1,\dots\,r) the $r$ gauge invariant moduli of the family,
(described as the vanishing locus of an appropriate polynomial)
one makes the identifications:
\begin{equation}
u_i~\to~\langle \, d_{\alpha_1\dots\alpha_{i+1}} Y^{\alpha_1}
\,\dots\, Y^{\alpha_{i+1}} \rangle \quad (i=1,\dots \, r)
\label{rie_1}
\end{equation}
where $Y^{\alpha}$ are the special coordinates of rigid special
geometry and $d_{\alpha_1\dots\alpha_{i+1}}$ is the restriction
to the Cartan subalgebra of the rank $i+1$ symmetric tensor
defining the $(i+1)$--th Casimir operator. The identification
(\ref{rie_1}) is only an asymptotic equality for large values
of $u_i$ and $Y^{\alpha}$; at finite values, the relation between
the moduli $u_i$ and the special coordinates
(namely the elementary fields appearing in the
lagrangian) is much more complicated. One considers the
derivatives :
\begin{equation}
\Omega_{u_i}\,{\stackrel{\rm def}{=}}\,
\partial_{u_i} \,\Omega\, =\,
\partial_{u_i}\left ( \matrix { Y^{\alpha}\cr
{\o{\partial {\cal F}}{\partial Y^{\alpha}}}\cr} \right)
\label{rie_2}
\end{equation}
where  $\Omega (u_i)$ is a section
of the flat $Sp(2r,\IR)$
holomorphic vector bundle  whose existence
is encoded in the definition of rigid special K\"ahler geometry
\cite{CDF,CDFVP,spec2}.
On one hand the K\"ahler metric is given by the following
general formula:
\begin{equation}
g_{i j^\star}~=~-\, {\rm i} \,
{\bar \Omega}^{T}_{{\bar u}_{j^\star}} \,\left (
\matrix {0&\bfone\cr -\bfone&0\cr} \right ) \, \Omega_{u_{i}}
\label{rie_3}
\end{equation}
On the other hand, one identifies the symplectic vectors
$\Omega_{u_{i}}$
with the period vectors:
\begin{equation}
 \Omega_{u_{i}}~=~\left ( \matrix {\int_{A^\alpha} \,
\omega^{i}\cr \int_{B^\alpha}\omega^{i}\cr} \right ) \quad
\quad (i=1,\dots\, r=\mbox{ genus})
\label{rie_3bis}
\end{equation}
of the $r$ holomorphic 1-forms $\omega^{i}$
along a canonical homology basis:
\begin{equation}
A_\alpha \cap A_\beta = 0 \quad \quad B_\alpha \cap B_\beta = 0
\quad \quad A_\alpha \cap B_\beta = -B_\beta \cap A_\alpha =
\delta_{\alpha\beta}
\label{rie_4}
\end{equation}
of a genus $r$ dynamical Riemann surface ${\cal M}_{1}[r]$.
The generic moduli space $M_r$ of genus $r$ surfaces is $3r-3$
 dimensional.
The dynamical Riemann surfaces ${\cal M}_1[r]$ fill an
$r$-dimensional sublocus $L_R[r]$.
The problem is that of characterizing intrinsically this locus.
Let
\begin{equation}
i \ : \ L_R[r] \rightarrow M_r
\end{equation}
be the inclusion map of the wanted locus and let
\begin{equation}
\label{hb}
{H}\stackrel{\pi}{\rightarrow} M_r
\end{equation}
be the Hodge bundle on $M_r$, that is the rank $r$ vector bundle whose
sections are the holomorphic forms on the Riemann surface
$\Sigma_r\in M_r$. As fibre metric on
this bundle one can take the imaginary part of the period matrix:
\begin{equation}
\label{imn}
{\rm Im}\, {\cal N}_{\alpha\beta} =\int_{\Sigma_r} \omega^{\alpha}\wedge
{\bar \omega}^{\beta^*}
\end{equation}
where $\omega^\alpha$ is a basis holomorphic one-forms. The locus $L_R[r]$
is defined by the following equation:
\begin{equation}
\label{boh}
i^*\partial \bar\partial ||\omega||^2=i^*{\cal K}
\end{equation}
where $||\omega||^2$ $=\int_{\Sigma_r}\omega\wedge{\bar\omega}$ is the norm
of any section of the Hodge bundle and ${\cal K}$ is the K\"ahler class
of $M_r$.
\par
Using very general techniques of algebraic geometry, the dependence
of the periods (\ref{rie_2}) on the moduli parameters can be
determined through the solutions of the Picard--Fuchs differential
system, once ${\cal M}_{1}[r]$ is  explicitly described
as the vanishing locus of a holomorphic superpotential
${\cal W}(Z,X,Y;u_i)$. In particular one can study the monodromy
group $\Gamma_M$ of the differential system
and the symmetry group of the potential
$\Gamma_{\cal W}$, that are related to the full group of
{\it electric--magnetic  duality rotations} $\Gamma_D$ as follows
\cite{LSW}:
\begin{equation}
\Gamma_{\cal W}~=~\Gamma_D/\Gamma_M
\label{rie_5}
\end{equation}
The elements of ${\Gamma_D}\supset\Gamma_M$ are given by integer
valued
symplectic matrices $\gamma \in Sp(2r,\ZZ)$
that act on the symplectic
section $\Omega$. Given the geometrical interpretation
(\ref{rie_3bis}) of these sections, the elements $\gamma \in
\Gamma_D \subset Sp(2r,\ZZ)$ correspond to changes of the
canonical homology basis respecting the intersection matrix
(\ref{rie_4}).
\par
To be specific we mention the results obtained for the gauge
groups ${G}=SU(r+1)$. The rank $r=1$ case, corresponding
to ${G}=SU(2)$, was studied by Seiberg and Witten in their
original paper \cite{SW1}. The extension to the general case,
with
particular attention devoted to the $SU(3)$ case, was
obtained in \cite{kltold,faraggi}. In all these cases the
dynamical Riemann
surface ${\cal M}_{1}[r]$ belongs to the
hyperelliptic locus of genus $r$ moduli space,
the general form of a hyperelliptic surface
being described (in inhomogeneous coordinates)
by the following algebraic equation:
\begin{equation}
w^2~=~P_{(2+2r)} (z)~=~\prod_{i=1}^{2 +2r} \, (z - \lambda_i)
\label{ipere}
\end{equation}
where $\lambda_i$ are the $2+2r$ roots of a degree $2+2r$
polynomial. The hyperelliptic locus
\begin{equation}
L_H [r]\subset M_r\quad ,\quad \mbox{dim}\,  L_H[r]=2r-1
\label{hlocus}
\end{equation}
is a closed submanifold of codimension
$r-2$ in the $3r-3$ dimensional moduli space of genus
$r$ Riemann surface\footnote{For genus 1, the moduli space is also
1--dimensional and the hyperelliptic locus is the full moduli
space.}. The $2r-1$ hyperelliptic moduli
are the $2r+2$ roots of the polynomial appearing in
(\ref{ipere}), minus three of them that
can be fixed at arbitrary points by means of fractional
linear transformations on the variable $z$.
Because of their definition, however, the dynamical Riemann
surfaces ${\cal M}_1[r]$, must have $r$ rather than
$2r-1$ moduli. We conclude that the $r$--parameter
family  ${\cal M}_{1}[r]$ fills a locus $L_R[r]$ of
codimension $r-1$ in the hyperelliptic locus:
\begin{equation}
L_R[r]\subset L_H [r],\hskip 0.4cm
\mbox{codim}\, L_R[r]=r-1, \hskip 0.4cm
\mbox{dim} L_R[r]=r .
\label{dynlocus}
\end{equation}
This fact is expressed by additional conditions imposed on
the form of the degree $2+2r$ polynomial of
eq.(\ref{ipere}). In references \cite{kltold,faraggi}
$P_{(2+2r)} (z)$ was determined to be of the following form:
\begin{eqnarray}
\label{squadra}
P_{(2+2r)} (z)&=& P_{(r+1)}^2 (z)  -
P_{(1)}^2(z)  \nonumber\\
~&=&\left ( P_{(r+1)}(z) + P_{(1)}(z) \right ) \,
\left ( P_{(r+1)}(z) - P_{(1)}(z) \right ) \,
\end{eqnarray}
where $P_{(r+1)}(z)$ and $P_{(1)}(z)$ are two polynomials
respectively of degree $r+1$ and $1$. Altogether we have
$r+ 3$ parameters that we can identify with the $r+1$
roots of $P_{(r+1)}(z)$ and with the two coefficients
of $P_{(1)}(z)$
\begin{equation}
\begin{tabular}{ll}
$P_{(r+1)}(z)~= \prod_{i=1}^{r+1} \, (z - \lambda_i)$
\hskip 0.3cm , \hskip 0.3cm
&$P_{(1)}(z)~=~\mu_1 \, z + \mu_0$ \ .
\end{tabular}
\end{equation}
Indeed, since the polynomial (\ref{squadra}) must be effectively
of order $ 2+2r$, the highest order coefficient of
$P_{(r+1)}(z)$ can be fixed to $1$ and the only independent
parameters contained in $P_{(r+1)}(z)$ are the roots. On the
other hand, since $P_{(1)}(z)$ contributes only subleading powers, both
of its coefficients $\mu_1$ and $\mu_0$ are effective
parameters. Then, if we take into
account fractional linear transformations,
three {\it gauge fixing} conditions can be imposed
on the $r+3$ parameters $\{ \lambda_i\},
\{\mu_i\}$. In ref. (\cite{kltold,faraggi}) this freedom was
used to set:
\begin{eqnarray}
\sum_{i=1}^{r+1} \, \lambda_i&=&0 \nonumber\\
\mu_1&=&0\nonumber\\
\mu_0&=&\Lambda^{r+1}
\label{lasceltadiwolf}
\end{eqnarray}
where $\Lambda$ is the dynamically generated  scale. With
this choice the $r$--parameter family of dynamical
Riemann surfaces is described by the equation:
\begin{equation}
w^2=\left ( z^{r+1} - \sum_{i=1}^{r} \, u_i \left (\lambda
\right )
\, z^{r-i} \right )^2 \, - \, \Lambda^{2r+2}
\label{wolf}
\end{equation}
where the coefficients
\begin{equation}
 u_i \left (\lambda_1, \dots ,\lambda_{r+1} \right )
\qquad\qquad (i=1,\dots \, r)
\end{equation}
are symmetric functions of the $r+1$ roots constrained by the first
of eq.s (\ref{lasceltadiwolf}) and can be identified with the
moduli parameters introduced in eq.(\ref{rie_1}). In the particular
case $r=1$, the gauge--fixing (\ref{lasceltadiwolf}) leads to
the following quartic form for the elliptic curve studied in \cite{SW1}
\begin{equation}
w^2=\left ( z^2 - u \right )^2 - \Lambda^{4}=
z^4 - 2\, u \, z^2 + u^2 - \Lambda^4
\label{inhomseiwit}
\end{equation}
Of course other gauge fixings give equivalent descriptions of
${\cal M}_1[r]$; however, for our next purposes, it is particularly
important to choose a
gauge fixing of the $SL(2,\IC)$ symmetry such that the equation
${\cal M}_1[r]$ can be recast in the form of a Fermat polynomial
in a weighted projective space deformed by the marginal operators of
its chiral ring. In this way it is quite easy to study the symmetry
group
of the potential $\Gamma_{\cal W}$
identifying the $R$-symmetry group and to
derive the explicit form of the Picard-Fuchs equations satisfied by
the
periods. This is relevant for the embedding of the monodromy
and $R$-symmetry groups in $Sp(2r,\ZZ)$. The alternative gauge-fixing
that we choose is the following:
\begin{eqnarray}
\sum_{i=1}^{r+1} \, \lambda_i&=&0 \nonumber\\
\mu_1 \, \mu_0 + \,
\left ( \sum_{i=1}^{r+1} \, {\o{1}{\lambda_i}} \right )
\prod_{i=1}^{r+1} \, \lambda_i^2 \,
&=&0\nonumber\\
-\mu_0^2 + \prod_{i=1}^{r+1} \, \lambda_i^2&=&1
\label{lamiascelta}
\end{eqnarray}
To appreciate the convenience of this choice
 let us consider the general inhomogeneous form of
the equation of the hyperelliptic surface \eqn{squadra}
and let us (quasi-)homogenize it by setting:
\begin{equation}
\begin{tabular}{ll}
$w={\o{ Z}{Y^{r+1}}}$&$z={\o{ X}{Y}}$ \ .
\end{tabular}
\label{omogeneizzazione}
\end{equation}
With this procedure (\ref{squadra}) becomes a quasi--homogeneous
polynomial constraint:
\begin{eqnarray}
0&=&{\cal W}\left ( Z,X,Y;\{\lambda\},\{ \mu \} \right )\nonumber\\
~&=& - \, Z^2 + \left ( \prod_{i=1}^{r+1} \,
\left ( X - \lambda_i  \, Y\right ) \right )^2 -
\left ( \mu_1 \, X \, Y^{r} + \mu_0 \, Y^{r+1} \right )^2
\label{homog_1}
\end{eqnarray}
of degree:
\begin{equation}
 \mbox{deg} \, {\cal W}~=~2r+2
\label{gradino}
\end{equation}
in a weighted projective space $W\IC\IP^{2;r+1,1,1}$, where
the quasi--homogeneous coordinates $Z$, $X$, $Y$ have degrees
$r+1$,$1$ and $1$, respectively. Adopting the notations of
\cite{kimetal}, namely denoting by\footnote{Note the difference of
notation: $W\IC\IP^{n;q_1,q_2,\dots,q_{n+1}}$ is the full weighted
projective space, in which \eqn{pesataequazione} is a hypersurface.}
\begin{equation}
WCP^n(d;q_1,q_2,\dots,q_{n+1})_{\chi}
\label{pesataequazione}
\end{equation}
the zero locus (with Euler number $\chi$)
of a quasi--homogeneous polynomial of degree $d$ in an
$n$--dimensional weighted projective space, whose  $n+1$
quasi--homogeneous coordinates have weights $q_1$,$\dots$,$q_{n+1}$:
\begin{equation}
{\cal W}\left ( \lambda^{q_1} \, X_1,\dots\,
\lambda^{q_{n+1}} \, X_{n+1}\right )~=~\lambda^d \,
{\cal W}\left ( X_1,\dots , X_{n+1} \right )
\quad \quad \forall \lambda \in \IC
\label{rie_8}
\end{equation}
we obtain the identification:
\begin{equation}
{\cal M}_{1}[r] ~=~WCP^2(2r+2;r+1,1,1)_{2(1-r)}
\label{agnizione}
\end{equation}
that yields, in particular:
\begin{equation}
{\cal M}_{1}[1] =WCP^2(4;2,1,1)_{0}\ ;\qquad
{\cal M}_{1}[2] =WCP^2(6;3,1,1)_{2}\ .
\label{cognizione}
\end{equation}
for the $SU(2)$ case studied in
\cite{SW1} and for
the $SU(3)$ case studied in \cite{kltold,faraggi}.
Using the alternative gauge fixing (\ref{lamiascelta}), the
quasi--homogeneous Landau--Ginzburg superpotential (\ref{homog_1}),
whose vanishing locus defines the dynamical Riemann surface, takes
the standard form of {\it  a Fermat superpotential} deformed
by the marginal operators of its {\it chiral ring}:
\begin{equation}
{\cal W}\left ( Z,X,Y;\{\lambda\},\{ \mu \} \right ) =
- Z^2 + X^{2r+2} + Y^{2r+2}
+\sum_{i=1}^{2r-1} \, v_i \left ( \lambda \right ) \,
X^{2r+1-i} \, Y^{i+1}
\label{fermatform}
\end{equation}
The coefficients $v_i \, \quad (i=1,\dots\, 2r-1)$ are the
$2r-1$ moduli of a hyperelliptic curve. In our case, however,
they are expressed as functions of the $r$ independent
roots $\lambda_i$ that remain free after the gauge--fixing
(\ref{lamiascelta}) is imposed.  The coefficients $v_i$
have a simple expression as symmetric functions of
the $r+1$ roots $\lambda_i$ subject to the constraint
that their sum should vanish:
\begin{eqnarray}
v_1 \left ( \lambda \right )&=& \sum_{i} \lambda_i^2 \, + \,
4 \, \sum_{i<j} \, \lambda_i \, \lambda_j\nonumber\\
v_2 \left ( \lambda \right )&=&-\, 2 \, \sum_{i<j} \,
\left ( \lambda_i^2\, \lambda_j \, +\, \lambda_i\, \lambda_j^2 \,
\right ) \, - \, 8 \, \sum_{i<j<k} \, \lambda_i \, \lambda_j \,
\lambda_k \nonumber\\
v_3 \left ( \lambda \right )&=& \sum_{i<j} \,
 \lambda_i^2\, \lambda_j^2 \, +\, 16 \,
\sum_{i<j<k} \,\left ( \lambda_i^2 \, \lambda_j \,
\lambda_k  \, + \, \lambda_i \, \lambda_j^2 \,
\lambda_k \, + \, \lambda_i \, \lambda_j \,
\lambda_k^2 \right )\nonumber\\
v_4 \left ( \lambda \right )&=& - 2\, \sum_{i<j<k} \,
 \left ( \lambda_i^2 \, \lambda_j^2 \,
\lambda_k \, + \lambda_i^2 \, \lambda_j \,
\lambda_k^2 \, + \lambda_i \, \lambda_j^2 \,
\lambda_k^2 \right )\nonumber\\
&~& -\, 8 \,
\sum_{i<j<k<\ell} \,\left ( \lambda_i\, \lambda_j \,
\lambda_k  \,\lambda_\ell^2 \, + \,
\lambda_i\, \lambda_j \,
\lambda_k^2 \,\lambda_\ell \, + \,
\lambda_i\, \lambda_j^2 \,
\lambda_k \,\lambda_\ell \, + \,
\lambda_i^2 \, \lambda_j\,
\lambda_k \,\lambda_\ell \,
\right )\nonumber\\
v_5 \left ( \lambda \right )&=&\dots\dots
\label{simmetrichefunctie}
\end{eqnarray}
In particular for the first two cases $r=1$ and $r=2$
we respectively obtain:
\begin{eqnarray}
{\cal M}_1[1] &\hookrightarrow ~&\nonumber\\
0&=&{\cal W}\left ( Z,X,Y;v=2u\right )~=~
-Z^2 + X^4 + Y^4 +  v (\lambda ) \, X^2 Y^2 \label{mamma}\\
&~&\lambda_1 + \lambda_2 =0\nonumber\\
&~&\mu_1=0 \nonumber\\
&~&\mu_0=\sqrt{\lambda_1^2 \lambda_2^2 - 1}\nonumber\\
v&=&\lambda_1^2 + \lambda_2^2 +4\lambda_1 \, \lambda_2 ~=~
-2\, \lambda_1^2\,
 {\stackrel{\rm def}{=}}\, 2 \, u
\label{seiwitequa}
\end{eqnarray}
\begin{eqnarray}
{\cal M}_1[2] &\hookrightarrow~&\nonumber\\
0&=&{\cal W}\left ( Z,X,Y;v_1, v_2, v_3\right )\nonumber\\
&=&- \, Z^2 \, +\,  X^6\,  +\,  Y^6 \,  + \,
v_1 \, X^4 Y^2\,  +\,  v_2 \,  X^3 Y^3\,
+\,  v_3 \, X^2 Y^4\nonumber\\
&~&\lambda_1 + \lambda_2 +\lambda_3=0\nonumber\\
&~&\mu_1=-{\o{\lambda_1 \, \lambda_2 \, \lambda_3}
{\sqrt{\lambda_1^2 \, \lambda_2^2\, \lambda_3^2 \, - \, 1}}}
\, \left ( \lambda_1 \, \lambda_2 + \lambda_1 \, \lambda_3 +
\lambda_2 \lambda_3 \right ) \nonumber\\
&~&\mu_0=\sqrt{\lambda_1^2 \lambda_2^2 \lambda_3^2 - 1}\nonumber\\
v_1 &=& 2 \, (\lambda_1  \lambda_2 + \lambda_1 \lambda_3 + \lambda_2
\lambda_3 )\nonumber\\
v_2 &=& -2 \, \lambda_1\, \lambda_2 \, \lambda_3\nonumber\\
v_3&=&-\mu_1^2 + (\lambda_1  \lambda_2 +
\lambda_1 \lambda_3 + \lambda_2 \lambda_3 )^2\nonumber\\
\label{lerchefermat}
\end{eqnarray}
Alternatively, using as independent parameters the coefficients
$u_i(\lambda)$ appearing in eq. (\ref{wolf}), we can characterize
the locus $L_R[r]$ of dynamical Riemann surfaces
by means of the following
equations on the hyperelliptic moduli $v_i$:
\begin{eqnarray}
\label{n2}
v_k &=& -2 u_k + \sum_{i+j = k -1} u_i u_j,\hskip 1cm k = 1,\ldots,r
\nonumber\\
v_{r+k} &=& \sum_{i+j = r+k-1}u_i u_j - \delta_{r-1,k} \mu_1^2
\end{eqnarray}
Considering now the Hodge filtration of the middle cohomology group
$H^{(1)}_{DR}({\cal M}_1[r])$:
\begin{eqnarray}
{\cal F}^{0} \, &\subset &\, {\cal F}^{1}\, \nonumber\\
{\cal F}^{0}~=~H^{(1,0)}\hskip 0.5cm &;&
\hskip 0.5cm
{\cal F}^{1}\equiv H^{(1)}_{DR} = H^{(1,0)} +
H^{(0,1)} \nonumber\\
\label{filtraggio}
\end{eqnarray}
the Griffiths residue map (\cite{griffith,fresoriabook})
provides an association between elements of ${\cal F}^{k}$
and  polynomials $P^{I}_{k|(2r+2)}(X)$
of the chiral ring ${\cal R} ({\cal W} )~{\stackrel{\rm def}{=}}~
 \IC [X]/\partial {\cal W}$ of degree $k|(2r+2) \equiv
(k+1)(2r+2)-r-3$
according to the following pattern:
\begin{equation}
\begin{tabular}{llll}
$\mbox{cohom.}$&$\mbox{deg}$&$\mbox{polynom.}$&$~$\\
$~$&$~$&$~$&$~$\\
${\cal F}^0$&$r-1$&$P^i_{0|(2r+2)}~$&$i=1,\ldots,r$\\
${\cal F}^1$&$3r+1$&$P^{i^*}_{1|(2r+2)}$&$i^*=1,\ldots,r$
\end{tabular}
\end{equation}
Explicitly, the periods of eq. (\ref{rie_3bis})
are represented by:
\begin{eqnarray}
\label{n3}
\int_C \omega^i &=& \int_C {X^{r-i} Y^{i -1}\over {\cal W}}\, \omega
\nonumber\\
\int_C \omega^{i^*} &=& \int_C {X^{r+i} Y^{2r- i + 1}\over {\cal W}^2}
\,\omega
\end{eqnarray}
where $C$ denotes any of the homology cycles and
$\omega=2 Z\, \dop X\wedge
\dop Y + X\,\dop Y\wedge \dop Z+ Y\,\dop Z\wedge \dop X$.
Using standard reduction techniques
\cite{picred,LSW}
one can obtain the first-order Picard-Fuchs differential system
\begin{equation}
\label{n4}
\left({\partial\over\partial v^I} \bfone - A_I(v)\right) V=0
\hskip 0.5cm I = 1,\ldots 2r-1
\end{equation}
satisfied by the $2r$-component vector:
\begin{equation}
\label{period}
V =
\left ( \matrix { \int_C \omega^i\cr
\int_C \omega^{i^*} \cr} \right ) \label{V2r}
\end{equation}
in the $2r-1$-dimensional moduli space of elliptic surfaces.
Using the explicit embedding of the locus $L_R[r]\subset L_H[r]$
described by equations (\ref{n2}), we obtain the Picard-Fuchs
differential system of rigid special geometry by a trivial
pull-back of eq. (\ref{n4}):
\begin{equation}
\label{n5}
\left({\partial\over\partial u^i} \bfone - A_I(v){\partial v^I\over
\partial u^i} \right) V=0.
\label{picf}
\end{equation}
The explicit solution of the Picard--Fuchs
equations  for $r=1,2$ has been given respectively in \cite{CDF,kltnew}.
The solution of the Picard--Fuchs equations for generic $r$ determines in
principle the period of the surface and the monodromy group. We do not attempt
to solve (\ref{picf}) for generic $r$, but rather we will determine in the
next section the monodromy group by relying on the defining polynomial of the
surface only.
\par
In the remaining part of this section we discuss the symmetry
group of $\cM_1(r)$, which, together with the monodromy group $\G_M$ defines
the duality group $\G_D$ according to equation (\ref{rie_5}).
This symmetry group can be defined by considering those linear
transformations ${\bf X} \, \to \, M_A{\bf X}$ of
the quasi--homogeneous
coordinate vector ${\bf X}=(X,Y,Z)$ such that
\begin{equation}
{\cal W}(M_A{\bf X};v)=f_A(v) {\cal W}({\bf X} ;\phi_A(v))\ ;\qquad
 \omega(M_A{\bf X} ) =   g_A(u)\omega({\bf X})
\label{diedro_2}
\end{equation}
where $\phi_A(v)$ is a (generally non--linear) transformation
of the moduli and $f_A(v)$ and $g_A(v)$ are compensating  overall
rescalings of the superpotential and the volume forms that depend
both on the moduli $v$
and on the chosen transformation $A$.
Let us restrict our attention to the sublocus (\ref{dynlocus}) of
the dynamical Riemann surface ${\cal W}({\bf X};u)=0$, so that the moduli
space geometry is a special Ka\"hler geometry with Ka\"hler potential
\begin{equation}
K = {\rm i} (Y^\alpha \bar\cF_\alpha-\bar Y^\alpha\cF_\alpha)\ .
\end{equation}
In this case, only the subgroup
$\Gamma_{\cal W}^0 \subset \Gamma_{\cal W}$
given by the transformations
that have a compensating rescaling factor for the symplectic section
(which is in the first line of \eqn{n3}
$f_A(u)/g_A(u)$),
acts as an isometry  group for the moduli space,
in contrast with curved
special geometry, where the whole $\Gamma_{\cal W}$ generates
isometries.
The hyperelliptic superpotential (\ref{fermatform}) admits a $
\Gamma_{\cal W}^0$ symmetry
group which is isomorphic to the dihedral group $D_{2r+2}$, defined
by the following relations on two generators $A,B$:
\begin{equation}
\label{n6}
A^{2r+2} = \bfone \hskip 0.5cm ; \hskip 0.5cm B^2 = \bfone
\hskip 0.5cm ; \hskip 0.5cm (AB)^2 = \bfone.
\end{equation}
The action of the generators on the moduli is the following.
Let $\alpha^{2r+2}=1$ be a $(2r+2)^{\rm th}$ root of the unit and
let the moduli $v_i$ be arranged into a column vector ${\bf v}$.
Then we have:
\begin{eqnarray}
\label{ab}
{\bf v}^\prime = A {\bf v}, \hskip 1cm
A = \left(\matrix{\alpha^2 & 0 & \ldots & 0 \cr 0 & \alpha^3 & \ldots & 0
\cr \vdots & \vdots & \ddots & \vdots \cr 0 & 0 & \ldots & \alpha^{2r}}
\right)\nonumber\\
{\bf v}^{\prime\prime} = B {\bf v}, \hskip 1cm
B = \left(\matrix{0 & 0 & \ldots & 1
\cr \vdots & \vdots & \ddots & \vdots
\cr 0 & 1 & \ldots & 0 \cr 1 & 0 & \ldots & 0}\right)
\end{eqnarray}
For the transformations $A$ and $B$ the compensating transformations
on the homogeneous coordinates $M_A$ and $M_B$ are
\begin{eqnarray}
M_A &=& \left ( \matrix{\alpha & 0 & 0 \cr 0 &1 &0 \cr 0& 0&1 }\right )
\hskip 0.3cm ; \hskip 0.3cm
M_B= \left ( \matrix{0 &1 &0 \cr  1 & 0 &0 \cr 0 &0 &1 }\right )
\end{eqnarray}
Consequently the differential Picard--Fuchs system for the period
(\ref{period}) of the
generic hyperelliptic surface has a $\Gamma_{\cal W}^0=D_{2r+2}$
symmetry as defined above
and the generators $A$ and $B$ act by means of
suitable $Sp(2 r, \ZZ)$ matrices on the period vector
(\ref{rie_3bis}). However the equations (\ref{picf})
are invariant only under the cyclic subgroup $\ZZ_{2r +2}
\subset D_{2r+2}$ generated by $A$.
Hence the potential $\tilde {\cal W}(u)={\cal W}\left(v(u)\right)$
of the $r$-dimensional locus
$L_R[r]$ of dynamical Riemann surfaces and the Picard-Fuchs first
order system admits only the duality symmetry  $\Gamma_{\tilde{\cal W}}^0=
\ZZ_{2r + 2}$.
\par
The physical interpretation of this group is R-symmetry.
Indeed, recalling eq. (\ref{rie_1}) we see that when each
of the elementary fields $Y^\alpha$ appearing in the lagrangian
is rescaled as $Y^\beta \to \alpha Y^\beta$, then the first $u_i$ moduli
are rescaled with the powers of $\alpha$ predicted by equation
(\ref{rie_1}).
According to the analysis of reference \cite{noi} this is precisely the
requested R-symmetry for the
topological twist. All the scalar components of the
vector multiplets have the same R-charge ($q_R=2$) under a
$U_R(1)$ symmetry of the classical action,
which is broken to a discrete subgroup
in the quantum theory. Henceforth the integer symplectic matrix
that realizes $A$ yields the R-symmetry matrix of rigid special
geometry for $SU(r+1)$ gauge theories.
An important problem is the derivation of the corresponding
R-symmetry matrix in $Sp(2r +4, \ZZ)$, when the gauge theory
is made locally supersymmetric by coupling
it to supergravity including also
the dilaton-axion vector multiplet suggested by string theory.
\section{ The monodromy group of $SU(r+1)$ rigid gauge theory}
   \label{ss:monoSUr1}
In this section we give a concise account of the explicit
construction of the monodromy group $\Gamma_M (r)\in Sp(2r;\ZZ)$
for the subclass of hyperelliptic surfaces given by
eqs. \eqn{squadra} or \eqn{wolf} associated with the $SU(r+1)$
rigid gauge theory. In our approach the required monodromy group is
selected as a particular subgroup of the monodromy group of a generic
hyperelliptic surface  of genus $r$
\begin{equation}
\Sigma_r \in L_H[r]:\{w^2=P_{(2r+2)}(z)
= z^{2r+2}+c_1\ z^{2r+1}+\cdots +c_{2r+1}\ z + c_{2r+2}=\Pi_{i=1}^{2r+2}
 (z-\lambda_i)\}\ ,
\label{poli}
\end{equation}
where only $2r-1$ of the $c_k$ are independent moduli.
The method presented here yields a complete solution
for any $SU(r+1)$ and uses some tools that were introduced in
\cite{CDR}.
Our results can be compared with those recently
obtained in \cite{kltnew}, where the particular case of $SU(3)$
has been thoroughly discussed and
where the corresponding periods of the theory have been
explicitly obtained\footnote{We refer the reader to that paper
for a more complete discussion of the geometrical
theory of monodromy and its relevance for $SU(n)$  $N=2$ gauge theories.
See also \cite{dansun} for the monodromy in $SO(2r+1)$ gauge theories}.
Our basic observation is the following:
the monodromy group for $\Sigma_r$
is given by a $2r$-dimensional representation   of $B(2r+2)$,
the braid group acting on $2r+2$ strands, on the
homology basis of $\Sigma_r$ . Indeed,
it is sufficient to recall that:\\
i)The monodromy group of a p-fold $\cM$ is given by
the representation, on
the homology basis of the p-fold, of the fundamental group
$\pi_1$ of the complement of the bifurcation set of $\cM$.\\
ii) For the case $\cM=\Sigma_r$, where $\Sigma_r$ is described by the
polynomial \eqn{poli}, denoting by $Q^{(2r-2)}$
the bifurcation set of eq. \eqn{poli}, and by
 $C$  the base point, we have
\begin{equation}
\pi_1({\bf CP}^{(2r-1)}-Q^{(2r-2)};C)\equiv B(2r+2)\ ,
\end{equation}
since $B(2r+2)$ is the fundamental group
of the space of polynomials of degree $2r+2$ with no multiple roots.
Indeed, the bifurcation set of a polynomial is given by the
submanifold in the moduli space $\{c_1,\ldots,c_{2r-1}\}$ where
two or more roots $\l_i$ coincide.
\par
The generators $t_i$ of $B(2r+2)$ correspond to the exchange of
the $i$-th and the $i+1$-th strand and satisfy the relations
\begin{eqnarray}
t_i t_{i+1} t_i &=& t_{i+1} t_i t_{i+1}\nonumber \\
t_i t_j &=& t_j t_i\qquad \qquad \qquad |i-j|\geq 2\ .
\label{brai1}
\end{eqnarray}
In particular, to each generator $t_i \in \pi_1\equiv B(2r+2)$
there corresponds a loop in the moduli space which exchanges the
roots $\l_i, \l_{i+1}$ of the polynomial and a vanishing
cycle of $\Sigma_r$. For a generic
hyperelliptic surface any two roots can be exchanged by a
suitable word in the generators $t_i$.
\par
Let us now consider the particular subclass of hyperelliptic surfaces
$\cM_1[r]\in L_R[r]$ described in the previous section, by
eqs. \eqn{squadra},\eqn{wolf}. Corresponding to the factorization
in eqs. \eqn{squadra},\eqn{wolf}
we have a natural splitting of the  $2r+2$ roots of $P_{(2r+2)}$ into
two sets
\begin{equation}
\{\l_1,\ldots,\l_{r+1}\}\ \ \
{\rm and}\ \ \ \{\l_{r+2},\ldots,\l_{2r+2}\}\ .
\label{brai3}
\end{equation}
It is obvious that for the particular surface $\cM_1[r]$
the fundamental group
$\pi_1$ mentioned above will be generated by those elements
$t_i\in B(2r+2)$ which respect the
splitting  \eqn{brai3}, that is
\begin{equation}
\{t_1,\ldots,t_{r},t^2_{r+1},
t_{r+2},\ldots,t_{2r+1}, T=(t_{1}t_{2}\cdots t_{2r+1})^{r+1}\}
\label{brai4}
\end{equation}
where $T$ corresponds to the exchange of the two sets of roots
($t_{1}\cdots t_{2r+1}$ corresponds to the cyclic permutation
$\{\l_1,\l_2,\ldots,\l_{2r+2}\}\to \{\l_2,\l_3,\ldots,\l_{2r+2},\l_1\}$).
We conclude that the fundamental group of the hyperelliptic surface
$\cM_1[r]$ is generated by the elements (\ref{brai4}).
The required monodromy
group $\Gamma_M[r]$ is therefore given by the representation on
the homology group
$H_1(\cM_1[r],\ZZ)$ of the generators (\ref{brai4}).
At this point the strategy for computing the explicit
monodromy of $\cM_1[r]$ is clear: one first obtains the monodromy group
of $\Sigma_r$ as a representation $M(t_i)$ of the $B(2r+2)$ generators on
the homology basis of $\cM_1[r]$. Then the monodromy group of $\cM_1[r]$
is given by the subgroup generated by
\begin{equation}
\left \{ M(t_1),\ldots,M(t_{r}), M^2(t_{r+1}),
 M(t_{r+2}),\ldots, M(t_{2r+1}), M(T)\right \}
\label{brai5}
\end{equation}
 \par
Let us then construct the monodromy group of
$\Sigma_r$ as a representation
of $B(2r+2)$ on $H_1(\Sigma_r;\ZZ)$. We first choose a basis of cycles
$(A^I,B_I)$ on the cut $z$-plane such that
\begin{equation}
A^I\cap A^J=B_I\cap B_J=0, \ A^I\cap B_J=-B_J\cap A^I=\d^I_J \quad
(I,J=1,\ldots,r)
\end{equation}
so that the homology intersection form $C$ takes the canonical form
\begin{equation}
C=\pmatrix{ 0 &\bfone_{r}\cr -\bfone_{r} & 0\cr}
\label{brai6}
\end{equation}
Actually we may take the cycles $A^I$ to encircle the couple of roots
$(\lambda_{2I}, \lambda_{2I+1})$, while the $B_I$ cycles encircle the set
of roots
$(\lambda_1,\lambda_2,\ldots,\lambda_{2I})$.
To a generic element $t\in B(2r+2)$ we may associate the corresponding
vanishing cycle $L$ of
$H_1(\Sigma_r,\ZZ)$, say
\begin{equation}
L=(n_{I}^e,n^{I}_m)
\end{equation}
where $(n_{I}^e ,n^{I}_m)$ are the components of $L$
with respect to the basis
$(A^I,B_I)$. Using the Picard--Lefschetz formula \cite{kltnew}
\begin{equation}
\d\to\d-(\d\cup L) L
\end{equation}
which represents  the transformation induced on the homology
by the vanishing cycle $L$ corresponding to the element $t \in
B(2r+2)$, it is easy to see that in the
given basis the corresponding monodromy
matrix $M(L)$ is given by:
\begin{equation}
M(L)=\bfone + L \otimes
(C L) \equiv\pmatrix{ \bfone+\vec n_e\otimes \vec n_m &
-\vec n_e\otimes \vec n_e\cr  \vec n_m\otimes \vec n_m &
\bfone -\vec n_m\otimes \vec n_e\cr}\ .
\end{equation}
Denoting by $L^{(i)}$ the homology element associated to $t_i$
($i=1,\ldots,2r+2$), their explicit form is found by imposing the braid
relations (\ref{brai1}) on $M(L^{(i)})$, which yield the constraints
\begin{eqnarray}
L^{(i)T}C L^{(j)} &=& 0\qquad\qquad \ |i-j|\geq 2 \nonumber\\
L^{(i)T} C L^{(i+1)}&=& 1\ .
\end{eqnarray}
The solution can be written as follows
\begin{eqnarray}
L^{(2j-1)} &=& (\vec e_j - \vec e_{j-1};\vec 0)\nonumber\\
L^{(2j)} &=& (\vec 0 ;-\vec e_j)\qquad\qquad j=1,\ldots,r \nonumber\\
L^{(2r+1)} &=& (-\vec e_r;\vec 0)\nonumber\\
L^{(2r+2)} &=& (\vec 0;\vec e_1+\cdots+\vec e_r)
\label{brai7}
\end{eqnarray}
where $\vec e_i$ is an orthonormal basis in $\IR^r$ ($\vec e_0=0$).
Notice that the electric charges of the odd-numbered $L$ and the
magnetic charges of the even-numbered $L$ are given in terms of the roots
and fundamental weights of $SU(r+1)$ \cite{kltnew}.
The restriction of the braid group generated by $M(L^{(i)})$ to
the subgroup given in (\ref{brai5}) (with ($M(L^{(i)})\equiv M(t_i)$)
 gives the monodromy group of the hyperelliptic curves $\cM_1(r)$ .
We stress that our construction selects uniquely the possible  entries
of $ L^{(i)}=(\vec n_e^{(i)},\vec n_m^{(i)})$,
corresponding to the values of the
electric and magnetic charges of any $SU(r+1)$ gauge theory.
\section{Coupling to Supergravity and the dynamical
\protect\\ Calabi-Yau manifold}\label{ss:SGCY}
When we couple vector multiplets to supergravity, in the scalar
sector {\it rigid special geometry} \cite{CDF}
is replaced by its local version, namely by {\it special geometry}
\cite{BDW,spec1,spec12,spec3,spec2,spec13}.
For the coupling of the microscopic N=2 gauge theory
we have two possibilities.
1) The most natural generalization
of the minimal coupling (\ref{mincoup_1}) is given by the
gravitational minimal coupling where the number of vector multiplets
$n={\rm dim}_{\IR} {\cal G}$ remains the same as in the rigid
theory and the scalar manifold is given by:
\begin{equation}
{\cal SK}_{local}(n) ~=~MK(n)~{\stackrel{\rm def}{=}} ~
{\o{SU(1,n)}{U(1)\times SU(n)}}
\label{mincoup_2}
\end{equation}
while the prepotential is:
\begin{eqnarray}
F(X)&=&(X^0)^2 - g_{IJ}^{(K)} \, X^I \,
X^J\nonumber\\
g_{IJ}^{(K)}&=&\mbox{positive definite Killing metric of
the Lie algebra}~{\bf G}\nonumber\\
X^\Lambda&=&(X^0,X^{I})~=~\mbox{local special coordinates}
\label{mincoup_3}
\end{eqnarray}
Even if such a choice is a consistent one from the supergravity point of
view, it is however not compatible with string theory,
since the multiplet containing the dilaton--axion is missing.
Furthermore, one can show \cite{noi}
that there is no off--shell defined R--symmetry that
can lead to an off-shell defined ghost number \cite{Damiano}
in the topological
twisted theory, and that in the topologically twisted theory,
 the moduli--space
of gravitational instantons has dimension $3\times \tau$
rather than $4 \times \tau$ ($\tau=$ Hirzebruch signature), as
needed to obtain non vanishing topological correlators of operators
associated with $0$--cycles and $2$--cycles of the four--manifold.
\par
These problems disappear if we consider instead
the generalization of the minimal coupling
selected by string theory, which, besides the $n={\rm dim}_{\IR}
{\cal G}$
 vector multiplets of the rigid theory,  requires also an
additional vector multiplet $\left ( A_\mu^S,\lambda^S_A , \,
\lambda^{S^\star A}, \, S, \, {\bar S} \right )$ containing
the dilaton--axion field:
\begin{eqnarray}
S&=&{\cal A} + {\rm i}\exp [D]\nonumber\\
\partial_{[\mu} \, B_{\nu\rho]}&=&\varepsilon_{\mu\nu\rho\sigma}
   \partial^\sigma {\cal A}
\label{dilatonaxion}
\end{eqnarray}
whose vacuum expectation value provides the effective gauge
coupling constant and theta--angle:
\begin{equation}
\langle \, S \, \rangle ~=~{\o{\theta}{2\pi}}\, + \,
{\rm i}{\o{1}{g^2}}
\label{thetangle}
\end{equation}
The tree level effective action is based on the
following homogeneous special manifold:
\begin{equation}
{\cal SK}_{local}(n+1) ~=~ST(n)~{\stackrel{\rm def}{=}} ~
{\o{SU(1,1)}{U(1)}}\, \otimes \, {\o{O(2,n)}{O(2)\times O(n)}}
\label{stmanif_1}
\end{equation}
that, according to a theorem proved sometime ago \cite{sertoi},
is also the
only special manifold admitting a direct product structure.
If we use  the coordinate frame of
Calabi--Visentini (\cite{calvis})
for the submanifold ${\o{O(2,n)}{O(2)\times O(n)}}$:
\begin{equation}
\label{calabvise}
\begin{array}{l}
X^{\Lambda}{\stackrel{\rm def}{=}}
\left( X^0 , X^1 , X^{I} \right)\\
X^{\Lambda} X^{\Sigma}\eta_{\Lambda\Sigma}=0\end{array}
\hskip 0.1cm :\hskip 0.5cm
\left\{\begin{array}{l}
X^0={1\over 2}\left(1+g_{IJ}^{(K)}Y^I Y^J \right)\\
\null \\
X^1={\rm i\over 2}\left(1-g_{IJ}^{(K)} Y^I Y^J \right)\\
\null \\
X^{I}=Y^I \quad \quad \{I=1,\dots ,\, n={\rm dim}_{\IR}{\cal G}\}
\end{array}\right.
\end{equation}
where
$g_{IJ}^{(K)}$ is the Killing metric of the Lie algebra $\bf G$
and $\eta_{\Lambda\Sigma}$ $={\rm diag} \left( +,+,-g_{IJ}^{(K)}\right)$,
the K\"ahler potential being (the full Ka\"hler potential still
contains a term $- \log\ i(\bar S-S)$),
\begin{equation}
{\cal K}(Y,{\bar Y})=  -
\log \left[ X^{\Lambda} {\bar X}^{\Sigma}\eta_{\Lambda\Sigma}\, \right]
= -\log \left[ {\o{1}{2}} \left ( 1 - 2
g_{IJ}^{(K)} Y^{I} \, {\bar Y}^{J} + | g_{IJ}^{(K)} Y^{I} Y^{J} |^2
\right ) \right],
\label{calabvise2}
\end{equation}
then the appropriate $4+2n$--dimensional symplectic section
determining special geometry is provided by\cite{CDFVP}:
\begin{equation}
\Omega(S,Y)~=~\left ( \matrix {X^\Lambda \cr F_\Sigma ~=~S \,
\eta_{\Sigma\Delta} \, X^\Delta }\right )
\label{stmanif_2}
\end{equation}
When the theory is classical and purely abelian, with matter fields carrying
no electric and magnetic charges, the supergravity based on the $ST(n)$
special manifold admits a continuous group of duality
transformations \'a la Gaillard-Zumino \cite{GZ}:
\begin{equation}
SL(2,\IR) \otimes O(2,n),
\label{stmanif_3}
\end{equation}
the symplectic embedding into $Sp(4+2n,\IR)$ being as follows.
\def\twomat#1#2#3#4{\left(\begin{array}{cc}#1& #2\\ #3 &
#4\end{array}\right)}
\begin{equation}
\label{so2nemb}
\begin{array}{ccc}
A\in O(2,n) &\hookrightarrow &\left(
\begin{array}{cc}A & {\bf 0} \\ {\bf
0} & \eta A \eta^{-1}
\end{array}\right)\in Sp(2n + 4,{\IR})\\ & & \\
\left(\begin{array}{cc}a & b\\c & d
\end{array}
\right) \in Sp(2n+4,
\IR)&\hookrightarrow &
\twomat{a\bfone}{b\eta^{-1}}{c\eta}{d\bfone}\in Sp(2n + 4,{\IR}) \ ,
\end{array}
\end{equation}
where $A^T\eta A=\eta$.
On the other hand, consider the abelian phase of a spontaneously broken
Yang--Mills theory coupled to supergravity. If one takes into account
the massive charged modes,
the duality group $\Gamma_D^{loc}$ is a discrete group.
The reason is that the lattice of electric
and magnetic charges of the BPS saturated states must be preserved by the
duality rotations.  This happens even in those cases where
the local special geometry of the moduli
space does not receive quantum corrections and
remains the same as that of $ST(n)$
[described by eq. (\ref{calabvise})]. In these cases the duality
group $\Gamma_D^{loc}$ is a discrete  subgroup of
(\ref{stmanif_3}):
\begin{equation}
\label{modif1}
\Gamma_D^{loc}\subset SL(2,\ZZ) \otimes O(2,n;\ZZ )\ ,
\end{equation}
the embedding into $Sp(2n + 4,\ZZ)$ being the restriction to the integers
of the embedding (\ref{so2nemb}).
\par
In general, however, the local geometry of the moduli space $ST(n)$ is
modified by  perturbative and  non perturbative effects.
Therefore, considering the  effective N=2 lagrangian describing
the dynamics of the massless modes, that admits the $r$--dimensional
maximal torus $ H\subset  G$ as gauge group,
we are faced with the problem of finding  the $r+1$--dimensional
special manifold ${\hat {ST}}(r)$
that encodes the complete structure of this lagrangian and
the exact quantum duality group ${\Gamma}_D^{local}$.
\par
We note that ${\hat {ST}}(r)$ is
a quantum deformation of the manifold $ST(r)$:
for large values of the moduli, namely in a
asymptotic region, to be appropriately defined, where the
quantum theory approaches its classical limit, the manifold
${\hat {ST}}(r)$ should reduce to $ST(r)$. This manifold is the
truncation to the Cartan--subalgebra fields of the manifold
$ST(r+\# \mbox{ of roots}=n)$, that corresponds to the {\it
gravitationally coupled microscopic gauge theory}. At the
same time, the quantum
duality group of the rigid theory ${\Gamma}^{rig.}_D$
should be embedded in the quantum duality group of the local theory
\begin{equation}
Sp(2r,\ZZ) ~\supset~{ \Gamma}^{rig.}_D~\hookrightarrow~
{\Gamma}^{loc.}_D ~\subset~Sp(4+2r,\ZZ)
\label{stmanif_7}
\end{equation}
In special coordinates
\begin{equation}
\begin{tabular}{lll}
$S={\o{{\hat X}^1}{{\hat X}^0}}$&$t^{i}={\o{{\hat X}^i}{{\hat X}^0}}$
&$i=2,\dots , r+1$
\end{tabular}
\label{stmanif_8}
\end{equation}
this means that
the prepotential of the {\sl quantum local special geometry} is
of the following form:
\begin{eqnarray}
{\cal F}^{loc}(S,t)~=~\left ( {\hat X}^0\right )^{-2}\,
F^{loc} \left ( {\hat X} \right )&=&
{\o{1}{r!}} \, S \, t^{i} \, t^{j}
\eta_{ij} + \Delta {\cal F}^{loc}(S,t)\nonumber\\
\lim_{ t^{i} \, \to \, t^{i}_0,\ S\,\to \, S_0}\,
\Delta {\cal F}^{loc}(S,t) & = & 0
\label{stmanif9}
\end{eqnarray}
the asymptotic region corresponding to a neighbourhood
of $S, t^{i} = S_0, t^{i}_0$ where $S_0, t^{i}_0$ are appropriate
values, possibly infinite. Eq.(\ref{stmanif9}) is the
local supersymmetry counterpart
 of eq.(\ref{mincoup_3}) that applies instead to
the case of rigid supersymmetry.
\par
The reason why we have put a hat on the $X^\Lambda$ is that
they cannot be directly identified with the $X^\Lambda$
introduced in eq.(\ref{calabvise}). Indeed, in the symplectic
basis defined by eqs. \eqn{calabvise},\eqn{stmanif_2}, namely
in the basis
where, according to the embedding
(\ref{so2nemb}), the $O(2,n)$ symmetry
and, hence, the gauge symmetry $ G \subset O(n) \subset
O(2,n)$ are linearly realized, the special geometry of the manifold
$ST(n)$ admits no description in terms of a prepotential
$F(X)~=~X^\Lambda F_\Lambda /2$. This is due to the constraint
$0=X^{\Lambda} \, X^{\Sigma} \, \eta_{\Lambda\Sigma}$ \cite{CDFVP}.
 Hence
although the Calabi--Visentini coordinates $Y^{I}$ are identified
with
the special coordinates of rigid special geometry, yet the
$X^\Lambda$ appearing in (\ref{calabvise}) and (\ref{stmanif_2})
are not independent special coordinates for local special geometry.
To obtain a prepotential one
needs to perform a symplectic rotation to a new basis:
\begin{eqnarray}
\left ( \matrix {{\hat X}^\Lambda \cr
\partial_\Sigma F({\hat X})\cr }\right )&=&M
\left ( \matrix {{ X}^\Lambda \cr
S \,\eta_{\Sigma\Delta}\,  X^\Delta\cr }\right )\nonumber\\
M&=&\left ( \matrix {
1 & 0 & -\bfone & 0 & 0 &{\bf 0} \cr
0 & 0 & {\bf 0}& 1 & 0 &\bfone\cr
{\bf 0} & -\bfone & {\bf 0} & {\bf 0} & {\bf 0} & {\bf 0}\cr
0 & 0 &{\bf 0} & {{1}\over{2}} & 0 &-{{1}\over{2}}\bfone\cr
-{{1}\over{2}} & 0 &-{{1}\over{2}}\bfone &0 & 0 &{\bf 0}\cr
{\bf 0} & {\bf 0} &{\bf 0}& {\bf 0} &-\bfone &{\bf 0}\cr
}\right) \, \in \, Sp(4+2n,\IR)
\label{stmanif_10}
\end{eqnarray}
leading to a new symplectic embedding:
\begin{equation}
\label{so2nembbis}
\begin{array}{ccc}
A\in O(2,n) &\hookrightarrow &M\, \left(
\begin{array}{cc}A & {\bf 0} \\ {\bf
0} & \eta A \eta^{-1}
\end{array}\right) \, M^{-1}\in Sp(2n + 4,{\IR})\\ & & \\
\left(\begin{array}{cc}a & b\\c & d
\end{array}
\right) \in Sp(2n+4,
\IR)&\hookrightarrow & M \,
\twomat{a\bfone}{b\eta^{-1}}{c\eta}{d\bfone} \, M^{-1}\in Sp(2n + 4,{\IR})
\end{array}
\end{equation}
After this change of basis the symmetric constant tensor $\eta_{ij}$
appearing in (\ref{stmanif9}) is not the positive
definite $g_{\alpha\beta}^{(K)}$, appearing in eq.(\ref{mincoup_3}),
namely the reduction to the Cartan--subalgebra  of the Killing metric
$g_{IJ}^{(K)}$.It is rather a form with
Lorentzian signature $(-,+,+$, $\dots , +)$.
Now, the basic idea to obtain the explicit form of the gravitationally
coupled  effective lagrangian is to identify  the special
K\"ahler manifold ${\hat {ST}}(r)$ with the  complex--structure
moduli space of  an  $r+1$--parameter family of
dynamical Calabi--Yau three--folds ${{\cal M}}_3 [r]$.
\par
This is the obvious
generalization of the procedure adopted in the rigid case. In
the  same way as the rigid special manifold
${\cal SK}^{rig}[r]$ is the moduli--space of an $r$--parameter
family of genus $r$ dynamical Riemann--surfaces ${\cal M}_1[r]$,
the local special manifold ${\hat {ST}}(r)$ is the moduli--space
of a family of Calabi--Yau threefolds. The relation between
local special geometry and the variations of Hodge--structures
of Calabi--Yau threefolds is well known \cite{spec3} but we have
of course to impose further requirements on ${\cal M}_3 [r]$ in
order for its moduli space to represent the gravitational coupling
of an already given rigid effective theory.
\par
Any N=2 globally supersymmetric field theory can be made locally
supersymmetric by coupling to N=2 supergravity. This is
always possible because of the off--shell
structure of N=2 supersymmetry. However
the procedure is generally one--to--many as a consequence of
the interplay between the auxiliary fields belonging to the
matter multiplets and those pertaining to the gravitational multiplet.
Once the latter are introduced we have an additional freedom
in framing the interaction and various results can be obtained
that would be the same if we had only the matter auxiliary fields
to play with. Correspondingly the infinite Planck--mass limit
\begin{equation}
M_P = {\o{1}{\kappa}} \, \to \, \infty
\label{maxplanck}
\end{equation}
of a locally supersymmetric theory is not the same thing as
a globally supersymmetric theory: this is a quite familiar
phenomenon in all the phenomenological applications of
supersymmetry.Therefore, in order to state which locally
supersymmetric theory can be regarded as the coupling of which rigid
theory, one needs some criteria.
\par
In the case of a rigid gauge
theory one uses its renormalizability to study the singularities
and monodromies produced at the perturbative level and then
guesses the
complementary singularities introduced by non perturbative effects.
This procedure is not available if we start from the gravitational
coupling of the microscopic gauge theory since this theory is no
longer renormalizable. Obviously one can calculate
perturbative effects
in string theory and then implement them in the effective lagrangian.
This is one possible route and corresponds to the gravitational
counterpart of the  procedure followed in the rigid case
\cite{AFGNT}. The now more difficult task of guessing the
complementary singularities remains and this amounts to guessing
a {\it dynamical Calabi--Yau} with the appropriate monodromies.
This argument shows that one can anyhow by--pass the string step and
go directly to the central question: namely
{\it which is the Calabi--Yau three--fold with the appropriate
monodromies?} Appropriate monodromies are those that include
the monodromies of the rigid theory. More specifically we
should have:
\begin{equation}
\begin{array}{ccccccc}
Sp(2r+4,\ZZ) & \supset &{ \Gamma}_D^{local} & \supset &
{\Gamma}_D^{rigid} &  & \\
{\Gamma}_D^{local}& \supset & \Gamma_M^{local} & \supset &
\Gamma_M^{rigid} & \subset &{\Gamma}_D^{rigid}
\end{array}
\label{cucco}
\end{equation}
Recalling the
fundamental relation (\ref{rie_5}) between  the group
of electric--magnetic duality rotations and
the monodromy group, we also have:
\begin{equation}
\Gamma_{\cal W}^{rig} \, \subset \Gamma_{\cal W}^{local}
\label{ciccius}
\end{equation}
In the previous section we have studied the general form of
$\Gamma_{\cal W}^{rig}$ for $SU(r+1)$ gauge theories showing that
it is $\ZZ_{2r+2}$ and that it coincides with the R--symmetry
group. It follows that the symmetry group of the gravitationally
coupled theory, namely of the dynamical Calabi--Yau threefold,
should conveniently embed the rigid R--symmetry group. This
is the same request formulated in \cite{noi} in order
to be able to define the topological twist of the quantum theory.
\par
These are the  basic
criteria that allow to identify the corresponding matter
coupled supergravity as the locally supersymmetric version
of the already determined globally supersymmetric effective gauge
theory.
\par
Let us summarize the results of our discussion.
The family of dynamical Calabi--Yau manifolds ${\cal M}_3 [r]$
must satisfy the following conditions:
\begin{itemize}
\item{ ${\cal M}_3 [r]$ must be an $r+1$--parameter family
of algebraic three--folds in a (product of ) weighted--projective
spaces described by the vanishing ${\cal W}_i=0 ~(i=1,\dots ,p)$ of
the $p$ addends of a Landau--Ginzburg superpotential:
\begin{equation}
{\cal W}(X^1,\dots\, X^m;\psi_1,\dots , \psi_r)~=~\sum_{i=1}^{p}
{\cal W}_i(X^1,\dots\, X^m;\psi_1,\dots , \psi_{r+1})
\label{lgsuperpotential}
\end{equation}
depending on the $r+1$--parameters $\psi_1,\dots ,\psi_{r+1}$
and on the $m=3+p+1$ quasi--homogeneous coordinates of the ambient
space.}
\item{ The first Chern class of the hypersurface family must
obviously vanish
\begin{equation}
c_1 \left ( {\cal M}_3 [r] \right ) ~=~0
\label{svanisciclasse}
\end{equation}
}
\item{ The family ${\cal M}_3 [r]$ must contain some
multiple cover of the
family ${\cal M}_1 [r]$ of genus $r$ Riemann surfaces.
This guarantees the embedding of the rigid R--symmetry
group $\ZZ_{2r+2}$  into the symmetry group $\Gamma_{\cal W}$
of the Calabi--Yau potential.
}
\item{ Writing the degree $\nu$ superpotential (\ref{lgsuperpotential})
as  the deformation of a reference superpotential ${\cal W}_0(X)$
\begin{equation}
{\cal W}(X;\psi)~=~{\cal W}_0(X)\, +
\, \sum_{I=0}^{r} \, \psi_I\, P^{I}_{1|\nu}
\label{deformatus}
\end{equation}
The chiral ring :
\begin{equation}
{\cal R}_{{\cal W}_0}~=~{\o{\IC [X]}{\partial {\cal W}_0}}
\label{chiranello}
\end{equation}
of the degree $\nu$ reference superpotential (\ref{lgsuperpotential})
 should  contain
a subring of dimension $2+2\times (r+1)$ spanned by polynomials
$P^{J}(X)$ of the following degrees:
\begin{equation}
\begin{tabular}{lll}
$\mbox{degree}$&$\mbox{polyn.}$&$\mbox{index range}$\\
$~$&$~$&$~$\\
$0\times\nu$&${\cal P}_{0|\nu}=1$&$~$\\
$~$&$~$&$~$\\
$1\times\nu$&${\cal P}^{0}_{1|\nu}(X)$&$~$\\
$1\times\nu$&${\cal P}^{\alpha}_{1|\nu}(X)$&$\alpha=1,\dots \, r$\\
$~$&$~$&$~$\\
$2\times\nu$&${\cal P}^{0}_{2|\nu}(X)$&$~$\\
$2\times\nu$&${\cal P}^\alpha_{2|\nu}(X)$&$\alpha=1,\dots \, r$\\
$~$&$~$&$~$\\
$3\times\nu$&${\cal P}^{top}_{3|\nu}(X)$&$~$\ .
\end{tabular}
\end{equation}
such that they satisfy the relations
\begin{equation}
\begin{tabular}{lll}
$~$&${\cal P}^{0}_{1|\nu}\, \cdot \, {\cal P}^{0}_{1|\nu} \,
\sim \, 0 $&$~$\\
${\cal P}^{0}_{1|\nu} \, \cdot \, {\cal P}^{\alpha}_{1|\nu} \,
\sim \, {\cal P}^{\alpha}_{2|\nu}$&
$~$&
${\cal P}^{\alpha}_{1|\nu} \, \cdot \, {\cal P}^{\beta}_{1|\nu} \,
\sim \,  g_{\alpha\beta}^{(K)} \, {\cal P}^{0}_{2|\nu}$\\
$~$&
${\cal P}^{0}_{1|\nu} \, \cdot \,{\cal P}^{\alpha}_{1|\nu} \, \cdot \,
{\cal P}^{\beta}_{1|\nu} \, \sim \, g_{\alpha\beta}^{(K)}\,
{\cal P}^{top}_{3|\nu}$&
$~$\ .
\end{tabular}
\label{subanello}
\end{equation}
This condition guarantees that in the asymptotic region where
the classical limit of the moduli space is attained,
the geometry of ${\hat {ST}}[r]$ does indeed converge to that
of $ST[r]$. The fusion coefficients of the chiral ring
displayed in eq. (\ref{subanello}) coincide with the anomalous
magnetic moments of the $ST[r]$ manifold in its asymptotic
region.}
\end{itemize}
An obvious approach to the construction of suitable
dynamical Calabi--Yau threefolds for
rank $r$ locally supersymmetric gauge theories
is that of identifying these manifolds with
the mirrors of Calabi--Yau threefolds with
$h^{(1,1)}=r+1$:
\begin{equation}
{\cal M}_3[r]~=~{\tilde M}_3\left ( h^{1,1}=r+1\, ;
\, h^{2,1}=x\right )
\label{specchiata}
\end{equation}
Next one looks at the duality--monodromy groups and at the
structure of their deformation ring to see whether the other
requests are satisfied. This programme corresponds to a viable
possibility if the class of manifolds with given $h^{(1,1)}=r+1$ is
known and small. Such a situation occurs, under
additional reasonable assumptions, for low values
of $r$, in particular for $r=1$ and $r=2$. Restricting
one's attention to those threefolds that are described
as the vanishing locus of a single polynomial constraint
in weighted $WCP^4$, the class of $h^{(1,1)}=2,3$
threefolds is known \cite{cand} and displayed below
\begin{equation}
\label{tabella}
\begin{array}{ccccc}
\quad &\mbox{Hypersurface}&\quad & h^{1,1} & h^{2,1} \\
\quad &\quad & \quad &\quad & \quad \\
\# 1 &WCP^4(8;2,2,2,1,1)&\quad  & 2 & 86 \\
\quad &\quad & \quad &\quad & \quad \\
\# 2 &WCP^4(12;6,2,2,1,1)&\quad  & 2 & 128 \\
\quad &\quad & \quad &\quad & \quad \\
\# 3 &WCP^4(12;4,3,2,2,1)&\quad  & 2 & 74 \\
\quad &\quad & \quad &\quad & \quad \\
\# 4 &WCP^4(14;7,2,2,2,1)&\quad  & 2 & 122 \\
\quad &\quad & \quad &\quad & \quad \\
\# 5 &WCP^4(18;9,6,1,1,1)&\quad  & 2 & 272 \\
\quad &\quad & \quad &\quad & \quad \\
\# 6 &WCP^4(12;6,3,1,1,1)&\quad  & 3 & 165 \\
\quad &\quad & \quad &\quad & \quad \\
\# 7 &WCP^4(12;3,3,3,2,1)&\quad  & 3 & 69 \\
\quad &\quad & \quad &\quad & \quad \\
\# 8 &WCP^4(15;5,3,3,3,1)&\quad  & 3 & 75 \\
\quad &\quad & \quad &\quad & \quad \\
\# 9 &WCP^4(18;9,3,3,2,1)&\quad  & 3 & 99 \\
\quad &\quad & \quad &\quad & \quad \\
\# 10 &WCP^4(24;12,8,2,1,1)&\quad  & 3 & 243 \\
\end{array}
\end{equation}
\par
Hence, under these assumptions, for the gravitational
coupling of an $r=1$ gauge theory, (i.e. for the $G=SU(2)$ case)
we have five possibilities distinguished by five
different values of the second Hodge number $h^{(2,1)}$.
Since this number counts
the K\"ahler classes of the mirror manifold under
consideration it has no relevance as long as we deal
with locally supersymmetric pure gauge theories.
So we are allowed to inquiry which of these manifolds
satisfy the additional embedding criteria outlined above.
\par
Consider, for instance the second model in table (\ref{tabella}).
 Its mirror manifold  with $h^{(1,1)} = 128$,
$h^{(2,1)} = 2$ is described as the vanishing locus of the following
weighted projective polynomial
\begin{equation}
\label{mirvafa}
{\tilde W} = Z_1^{12} + Z_2^{12} + Z_3^6 + Z_4^6 + Z_5^2
-12 \psi Z_1 Z_2 Z_3 Z_4 Z_5 - 2\phi Z_1^6 Z_2^6
\end{equation}
This two moduli potential admits the ${\Gamma}_{\cal W}=\ZZ_{12}$
symmetry given by:
\begin{equation}
\label{z12}
\left(\begin{array}{c} \psi \\ \phi\end{array}\right)
\rightarrow \twomat {\alpha^{11}}{0}{0}{\alpha^6}
\left(\begin{array}{c} \psi \\ \phi\end{array}\right)
\end{equation}
where $\alpha$ denotes a $12^{\rm th}$ root of the unity.
Clearly $\ZZ_{12}$ contains a subgroup $\ZZ_4$ acting as
\begin{equation}
\label{z12bis}
\left(\begin{array}{c} \psi \\ \phi\end{array}\right)
\rightarrow \twomat {\alpha^{\prime 3}}{0}{0}{\alpha^{\prime 2}}
\left(\begin{array}{c} \psi \\ \phi\end{array}\right)
\end{equation}
with $\alpha^{\prime 4}=1$.
This $\ZZ_4$ group should be the R-symmetry group
of the rigid SU(2) theory
which, therefore, should be embedded in the
gravitational symplectic group
$Sp(6,\ZZ)$ with generators  $A = (A_{12})^3$
where $A_{12}$ is the matrix
generating the $\ZZ_{12}$ $\Gamma_{\cal W}$ group in $Sp(6,\ZZ)$.
Such a triple covering  of the rigid theory R-symmetry
inside the gravitational one (and quite possibly also of the
 monodromy group)
 appears to be the
result of a triple covering (apart from exceptional points) of a
dynamical Riemann\footnote{It is important to stress that we do
not mean that such Riemann surface should be identified with the
rigid theory solution, but as a mathematical explanation why the
R--symmetry is $\ZZ_8$ rather than $\ZZ_4$. We expect that a  more
profound argument should be found in the microscopic original theory
in terms of space--time instanton sums.}
surface ${\cal M}_1[1]$ inside this
particular ${\cal M}_3[1]$.
To see this it suffices to set $Z_3 = Z_4 = 0$, $Z_1^3 = X,
Z_2^3=Y, Z_5 = Z$ in eq. (\ref{mirvafa}) and compare with
eq. (\ref{mamma}).
What is only a plausible conjecture for model $\# 2$
can instead be proved for model $\#1$ of table (\ref{tabella})
thanks to the explicit results contained in \cite{cand}.
Indeed the mirror
manifold of $WCP^4(8;2,2,2,1,1)$ has been studied in detail in
\cite{cand} and it is described as the vanishing
locus of the following octic superpotential\footnote{
Note that this example is connected through a conifold transition
\cite{GMS} to the Calabi--Yau manifold described by a quintic equation
in $\IC P_4$ ($h_{11}=1$, $h_{21}=101$).}:
\begin{equation}
{\cal W}=X_1^8 + X_2^8 + X_3^4 + X_4^4+ X_5^4 -8\psi \, X_1 X_2 X_3 X_4 X_5
-2\phi \, X_1^4 X_2^4
\label{octica}
\end{equation}
Also this manifold embeds (apart for exceptional points) a
multiple covering of the rigid theory elliptic surface ${\cal M}_1[1]$,
which, this time, is double rather than triple. For its realization
 it suffices to set, in eq. (\ref{octica}):
\begin{equation}
X_4=X_5=0 \quad\quad  X_1^2=X \quad\quad X_2^2=Y \quad\quad X_3^2=Z
\label{inletto}
\end{equation}
The potential (\ref{octica}) has a $\Gamma_{\cal W}=\ZZ_8$ symmetry
group whose action on the moduli $\psi,\phi$ is the following:
\begin{equation}
{\cal A} \, : \left \{ \psi \, , \, \phi\right \} \,
\longrightarrow \,
\left \{ \,\alpha \, \psi \, , \,- \phi\right \}\quad \quad
\alpha^8=1
\label{zetaotto}
\end{equation}
Clearly the transmutation of the rigid $\ZZ_4$ R-symmetry into
$\ZZ_8$ is due to the double covering, just as in the other
possible case
$WCP^4(12;6,2,2,1,1)$ of gravitational coupling, its transmutation
into $\ZZ_{12}$ was due to the triple covering. In the present case,
however, using the results of  \cite{cand}, this
statement can be verified explicitly. The integer symplectic
matrix that represents the $\ZZ_8$ generator on the periods has
been calculated in  \cite{cand} and has the following form:
\begin{equation}
Sp(6,\ZZ) \, \ni \, {\cal A} \, = \, \left (
\matrix{ -1 & 0 & 1 & -2 & 2 & 0 \cr -2 & 1 & 0 & -2 & 4 & 4\cr
0 & 1 & -1 & 0 & 0 & 2 \cr
1 & 0 & 0 & 1 & 0 & 0 \cr
  -1 & 0 & 0 & -1 & 1 & 1 \cr 1 & 0 & 0 & 1 & 0 & -1 \cr  }
\right )
\label{apotto}
\end{equation}
It is obtained by a change of basis which makes it
integer symplectic from the matrix given in \cite{cand}.
Its second power
\begin{equation}
{\cal R}_4={\cal A}^2~=~
\left (
\matrix{ -3 & 1 & -2 & -2 & 0 & 4 \cr -2 & 1 & -2 & 0 & 4 & 4 \cr 0
   & 0 & 1 & 0 & 4 & 0 \cr 0 & 0 & 1 & -1 & 2 & 0 \cr 0 & 0 & -1 & 1
   & -1 & 0 \cr -1 & 0 & 1 & -2 & 2 & 1 \cr  }
\right )
\label{erre4}
\end{equation}
is the generator of the $\ZZ_4$ R--symmetry of the original
theory. If we calculate its eigenvalues we find:
\begin{equation}
\mbox{eigenvalues of}\ \   {\cal R}_4~=~
\{ -1,i,-i,-1,i,-i\}
\label{autovalori}
\end{equation}
As we see,
in agreement with the properties of R--symmetry
discussed in \cite{noi}, (apart from an overall change of sign)
 there is a pair of complex conjugate eigenvalues $\pm i$
corresponding to the graviphoton and gravidilaton directions
and a unit eigenvalue corresponding to the physical vector
multiplet of $SU(2)$. Indeed
going to the basis of the eigenvectors of ${\cal R}_2 = {\cal R}_4^2$
we find
\begin{equation}
\tilde{\cal R}_2~=~\left (
\matrix{ -1 & 0 & 0 & 0 & 0 & 0 \cr 0 & -1 & 0 & 0 & 0 & 0
   \cr 0 & 0 & -1 & 0 & 0 & 0 \cr 0 & 0 & 0 & -1 & 0 & 0 \cr
  0 & 0 & 0 & 0 & 1 & 0 \cr 0 & 0 & 0 & 0 & 0 & 1 \cr  }
\right )
\label{olderre}
\end{equation}
This is the matrix that realizes the $\ZZ_2$ R--symmetry in a
Calabi--Visentini basis for the classical manifold
$ST(1)=SU(1,1)/U(1) \times O(2,1)/O(2)$. Hence, as we see,
the $\ZZ_2$ R--symmetry of the $SU(2)$ theory is indeed transplanted
into the gravitationally coupled theory and can be reduced to
the canonical form it takes as discrete subgroup of the $O(2)$
group in the corresponding classical moduli manifold, by means
of a change of basis. This change of basis, however,  is not
symplectic and in the same basis the monodromy matrices are not
symplectic integer valued. The quantum basis where both the R--symmetry
and the monodromies are symplectic integers is determined via
the Picard--Fuchs equations and gives for ${\cal R}_2$ the
expression
\begin{equation}
{\cal R}_2={\cal A}^4=
\left ( \matrix{ 3 & -2 & 4 & 0 & 0 & -4 \cr 0 & -1 & 0 & 0 & 0 & 0
\cr 0 &
  0 & -3 & 4 & 0 & 0 \cr 0 & 0 & -2 & 3 & 0 & 0 \cr 0 & 0 & 1 & -2
   & -1 & 0 \cr 2 & -1 & 0 & 4 & 0 & -3 \cr  }
\right )
\label{quantumRsym}
\end{equation}
The matrix ${\cal R}_2$ realizes, in the gravitational coupled theory,
the symmetry:
\begin{equation}
u \, \to \, - u
\label{umenou}
\end{equation}
of the rigid theory discussed below in eq.(\ref{diedro_5}).
\par
Next we verify that the deformation ring has the correct
structure.
The Griffith residue mapping associates the Hodge
filtration of the middle cohomology group:
\begin{eqnarray}
&{\cal F}^{0} \, \subset \, {\cal F}^{1}\, \subset \, {\cal F}^{2}
\, \subset \, {\cal F}^{3}&\nonumber\\
&F^{k}~=~H^{(3,0)}\,  +\,  H^{(2,1)}\,+ \dots +\, H^{(3-k,k)}&
\nonumber\\
&H^{(3)}_{DR}~=~~H^{(3,0)}\,  +\,  H^{(2,1)}\,+ \,  H^{(1,2)}\, +
\, H^{(0,3)}&
\label{wcp2_3}
\end{eqnarray}
with polynomials
\begin{equation}
{\cal P}^{\alpha}_{k|8}(X_1,X_2,X_3,X_4,X_5) \, \in \, {
\o{\IC [X]}{\partial {\cal W}^0}}
\label{wcp2_4}
\end{equation}
 of degrees $0,\, 8,\, 16,\, 24$, according to the following
pattern:
\begin{equation}
\begin{tabular}{llll}
$\mbox{cohom.}$&$\mbox{deg}$&$\mbox{polynom.}$&$~$\\
$~$&$~$&$~$&$~$\\
${\cal F}^0$&$0$&${\cal P}_{0|8}~=$&$1$\\
${\cal F}^1$&$8$&${\cal P}^\alpha_{1|8}$&$~$\\
${\cal F}^2$&$16$&${\cal P}^\alpha_{2|8}$&$~$\\
${\cal F}^3$&$24$&${\cal P}_{3|8}~=$&$X_1^6 \, X_2^6 \, X_3^6 \, X_4^{2}\,
 X_5^2~
{\stackrel{\rm def}{=}}~ {\cal P}^{top}$\ .
\end{tabular}
\end{equation}
where the deformations should satisfy the following algebra:
\begin{equation}
\begin{tabular}{l}
${\cal P}^{0}_{1|8}\, \cdot \, {\cal P}^{0}_{1|8} \,
\sim \, 0 $\\
${\cal P}^{0}_{1|8} \, \cdot \, {\cal P}^{0}_{2|8} \, \sim \,
{\cal P}^{top}$\\
${\cal P}^{1}_{1|8} \, \cdot \, {\cal P}^{1}_{2|8} \, \sim \,
{\cal P}^{top}$\\
${\cal P}^{0}_{1|8} \, \cdot \,{\cal P}^{1}_{1|8} \, \cdot \,
{\cal P}^{1}_{1|8} \, \sim \, {\cal P}^{top}$\ .
\end{tabular}
\label{wcp2_6}
\end{equation}
The deformation ${\cal P}^{1}_{1|8}$ corresponds to the
matter multiplet while the deformation ${\cal P}^{0}_{1|8}$
corresponds to the additional dilaton--axion multiplet  and
the algebra (\ref{wcp2_6}) guarantees
that the classical limit of the moduli--space (obtained for
large complex structures $\psi_I \, \to \, \infty$ ) is given by
the coset manifold
\begin{equation}
ST(1)~=~{\o{SU(1,1)}{U(1)}}\otimes {\o{O(2,1)}{O(2)}}
\label{wcp2_8}
\end{equation}
as requested by tree--level string theory. From the explicit
identification of the deformation polynomials:
\begin{eqnarray}
 {\cal P}_{0|8}&=&1\nonumber\\
 \cP^{0}_{1|8}&=&X_1^4 \, X_2^{4}\nonumber\\
\cP^{1}_{1|8}&=& X_1  \, X_2  \, X_3  \, X_4 \,  X_5 \nonumber\\
\cP^{0}_{2|8}&=&X_1^2 \, X_2^2 \, X_3^2 \, X_4^2 \, X_5^2\nonumber\\
\cP^{1}_{2|8}&=&X_1^3 \, X_2^3 \, X_3 \, X_4 \, X_5 \nonumber\\
\cP_{3|8}&=&X_1^4 \, X_2^4 \, X_3^2 \, X_4^2 \, X_5^2
\label{wcp2_9}
\end{eqnarray}
we immediately obtain that (\ref{wcp2_6}) is a viable
candidate for the description of the moduli
space of a locally supersymmetric N=2 theory.
\par
The list of Calabi--Yau threefolds obtained in \cite{kava}
as examples of dual heterotic/type-II models where the matching of vector
and hypermultiplet numbers is realized, overlaps with the models
 selected by
our embedding criterion, as we have already emphasized in the
introduction. Moreover the check of the large $S$--limit of the
anomalous magnetic couplings $W_{ijk}$ made by these authors agrees
with our previous discussion.
\def\dop{{\rm d}}
\def\ii{{\rm i}}
\def\ee#1{{\rm e}^{#1}}
\section{Central charges and BPS states from three-form cohomology}
The purpose of this section is to describe BPS states and central-charge
formulas directly from the three-form cohomology of
Calabi--Yau threefolds. Strictly speaking it is only in
type II B models that the vector multiplets are associated
with three--forms since in type II A models they are rather
associated with the two--forms. Yet, by using mirror symmetry
we can always interchange the  moduli--space of K\"ahler
structures of one Calabi--Yau manifold (two--form case)
with the moduli--space of complex structures (three--form
case) of its mirror. Hence for definiteness we always refer
to the type II B case and to the three--form cohomology.
\par
It then will appear evident that,
under the assumption that Calabi--Yau classical
moduli space of three-form cohomology (in type-II theories)
describes the
quantum moduli space of vector multiplets in heterotic strings
(second-quantized mirror symmetry \cite{FHSV}),
the Calabi--Yau lattice of
saturated states will correspond to the lattice of
monopoles and dyons of the
heterotic quantum theory. In particular conifold points on Calabi--Yau,
as shown in \cite{stromco,GMS}
will correspond to monopole point (non-perturbative)
singularities in N=2 heterotic strings.\par
In order to carry out this program we will make use of the cohomology
decomposition of the self-dual five
form ${\cal F}$ (which exists in type-II
strings), adopting the results of \cite{FBC},
and the recent analysis of conifold points,
corresponding to vanishing three-cycles, as points at which
some hypermultiplets, carrying Ramond-Ramond
magnetic and electric charges, become massless.\par
The five-form of type-IIB theory is selfdual: ${\cal F} = \null^*
{\cal F}$, so that it satisfies both Bianchi identities and equations of
motions:
\begin{eqnarray}
\label{cc1}
\dop {\cal F} & = & 0\nonumber\\
\dop \null^*{\cal F}& = & 0.
\end{eqnarray}
When the ten-dimensional space-time is compactified to $M_4\times
{\cal M}_3$, where $M_4$ is four-dimensional space-time and ${\cal M}_3$
is a Calabi--Yau threefold, the Poincar\'e dual of the
five-form ${\cal F}$, which is a
five-cycle, can be decomposed along a basis $S_2^i\times C_3^\Lambda$,
where $S_2^i$ are two-cycles of $M_4$ ($i=1,\ldots,b^2(M_4)$) and
$C_3^\Lambda$ are three-cycles of ${\cal M}_3$.  Choosing a
symplectic basis
$(A_\Lambda,B^\Lambda)$, ($\Lambda = 0,1,\ldots h^{2,1}$) for the
three-cycles and introducing the dual basis
$(\alpha_\Lambda,\beta^\Lambda)$ of harmonic three-forms we can write:
\begin{equation}
\label{cc2}
{\cal F} = {\cal F}^\Lambda\alpha_\Lambda +
{\cal G}_\Lambda \beta^\Lambda
\end{equation}
where
\begin{equation}
\label{cc3}
{\cal G}_\Lambda =
{\rm Im}\, {\cal N}_{\Lambda\Sigma} {\tilde {\cal F}}^\Sigma +
{\rm Re}\, {\cal N}_{\Lambda\Sigma} {\cal F}^\Sigma.
\end{equation}
${\cal F}^\Lambda={\cal F}^\Lambda_{\mu\nu}
\dop x^\mu\wedge\dop x^\nu$ and
${\cal G}^\Lambda = {\cal G}^\Lambda_{\mu\nu}\dop x^\mu\wedge\dop x^\nu$
are respectively the electric and magnetic field strengths of the gauge
vectors emerging from ${\cal F}$ in the compactification.
When monopole and dyon states are present the topology of
space-time is modified, as it is well-known.
There are non-contractible spheres $S_2^i$
that surround the i$^{\rm th}$ singularity of the gauge
field corresponding to each monopole (or dyon) state and the
integrals of ${\cal F}^\Lambda$ or ${\cal G}^\Lambda$ on such
spheres yield the value of electric ($n^e_\Lambda$)
or magnetic ($n^\Lambda_m$) charges for the state
wrapped by $S_2$. For any such sphere we can write:
\begin{eqnarray}
\label{cc4}
\int_{S_2\times A_\Lambda}{\cal F} = \int_{S_2} {\cal G}_\Lambda & =
& n^e_\Lambda \nonumber\\
\int_{S_2\times B^\Lambda}{\cal F} = \int_{S_2} {\cal F}^\Lambda & =
& n^\Lambda_m
\end{eqnarray}
These are the integral charges, with respect to the integral
cohomology basis for $H^3$.
\par
However the physical charges, related to the ``central charge'' and the
$h^{2,1}$ complex charges (electric and magnetic)
are those associated with the
graviphoton $T^-_{\mu\nu}$ and with the others
vectors in the theory, as they appear in the transformation
laws of the gauginos\footnote{We write here the
transformations for left-handed fermions; by ${\cal F}^-$ we intend the
antiselfdual part of ${\cal F}$; we use the notation in which the index
$i = 1,\ldots h^{2,1}$ carried by the $h^{2,1}$ gauginos $\lambda^{iA}$
is extended to the range $\Lambda =
0,1,\ldots,h^{2,1}$ by writing $\lambda^{\Lambda A} = f^\Lambda_i
\lambda^{iA}$, $i=1,\ldots,h^{2,1}$ where $f^\Lambda_i =
D_i (\ee{{\cal K}\over 2} X^\Lambda)$ (for notations see \cite{CDFVP}).}:
\begin{equation}
\label{cc5}
\delta\lambda^{\Lambda A} = {\hat {\cal F}}^{\Lambda-}_{\mu\nu}
\gamma^{\mu\nu} \epsilon^{AB} \varepsilon_B
\hskip 1.5cm (A,B = 1,2).
\end{equation}
The graviphoton $T^-_{\mu\nu}$ appears in the gravitino transformation:
\begin{equation}
\label{cc6}
\delta\psi_{\mu A} = T^-_{\rho\sigma}\gamma^{\rho\sigma} \gamma_\mu
\epsilon_{AB} \varepsilon^B +\ldots ;
\end{equation}
Note that
\begin{equation}
\label{cc7}
T^-_{\mu\nu} = T_\Lambda {\cal F}^{-\Lambda}_{\mu\nu}=
(F_\Lambda {\cal F}^{-\Lambda}_{\mu\nu} -
X^\Lambda {\cal G}^-_{\Lambda\mu\nu})\ee{{\cal K}\over 2}
\end{equation}
where we used the graviphoton projector
\begin{equation}
\label{cc8}
T_\Lambda = (F_\Lambda -{\bar {\cal N}}_{\Lambda\Sigma}
X^\Sigma) \, \ee{{\cal K}\over{ 2}}
\end{equation}
and ${\hat{\cal F}}^{-\Lambda} T_\Lambda = 0$, i.e.
\begin{equation}
\label{cc9}
{\hat{\cal F}}^{-\Lambda} = {\cal F}^{-\Lambda} -\ii T^- {\bar X}^\Lambda
\ee{{\cal K}\over{ 2}}\hskip 1cm\left(T_\Lambda{\bar X}^\Lambda
\ee{{\cal K}\over{ 2}}
= -\ii\right)
\end{equation}
By the definition of $T^-_{\mu\nu}$ and the definition of the
holomorphic three-form on ${\cal M}_3$
\begin{equation}
\Omega(\phi) = X^\Lambda(\phi) \alpha_\Lambda -
F_\Lambda(\phi) \beta^\Lambda
\end{equation}
(we work in arbitrary coordinates $\phi$, so that
$X^\Lambda=X^\Lambda(\phi)$)
it follows (using also ${\bar F}_\Lambda {\cal F}^{-\Lambda}
- {\bar X}^\Lambda {\cal G}_\Lambda^- = 0$) that
\begin{equation}
\label{cc10}
\int_{S_2\times {\cal M}_3} {\cal F}\wedge\Omega = \int_{S_2}
\ee{-{{\cal K}\over 2}} T^- = (X^\Lambda n^e_\Lambda -
F_\Lambda n^\Lambda_m)\equiv Z(\phi)
\end{equation}
which is precisely the (holomorphic) central charge.
\par
The other charges,  for the $h^{2,1}$ electric
and magnetic field strengths, are
\begin{equation}
\label{cc11}
\int_{S_2\times {\cal M}_3} {\cal F}\wedge D_i\Omega = \int_{S_2}
{\rm Im}\, {\cal
N}_{\Lambda\Sigma} {\hat{\cal F}}^\Sigma D_i X^\Lambda = D_i Z
\equiv q_i(\phi)\end{equation}
In the above discussion we used the decomposition of the
five-form ${\cal F}$\cite{FBC}:
\begin{equation}
\label{cc12}
{\cal F} = {\cal F}^\Lambda \alpha_\Lambda - {\cal G}_\Lambda
\beta^\Lambda =
\ee{{\cal K}\over 2} (T^- {\bar \Omega} + T^+ \Omega + {\cal F}^{-i}
D_i\Omega + {\cal F}^{+i^*} D_{i^*}{\bar \Omega})
\end{equation}
recalling also the relation between ${\hat{\cal F}}^{-\Sigma}$ and ${\cal
F}^{-\Sigma}$ given by eq. (\ref{cc9}).
Now it is easy to prove that
\begin{equation}
\label{cc13}
({\cal N}_{\Lambda\Delta} - {\bar{\cal N}}_{\Lambda\Delta})
(\delta^\Lambda_\Sigma -\ii T_\Sigma {\bar X}^\Lambda) {\cal F}^{-\Sigma}
D_{i^*}{\bar X}^\Delta
\end{equation}
is the correct ansatz for the additional field strengths,
orthogonal to the
graviphoton, since, due to
\begin{equation}
\label{cc14}
({\cal N}_{\Lambda\Delta} - {\bar{\cal N}}_{\Lambda\Delta})
{\bar X}^\Lambda D_{i^*} {\bar X}^\Delta = 0
\end{equation}
the components of (\ref{cc13}) along $T^-_{\mu\nu}$ is zero. We have then
\begin{equation}
\label{cc15}
{\cal F}^{-i} = \ee{{\cal K}/2} g^{ij^*} D_{j^*} {\bar X}^\Lambda
({\cal N}_{\Lambda\Delta} - {\bar{\cal N}}_{\Lambda\Delta}) {\hat{\cal
F}}^{-\Delta}
\end{equation}
Therefore the difference between the quantized electric and
magnetic charges $n_\Lambda^e , n_m^\Lambda$ and the moduli dependent
charges $(Z,q_i)$ is that the first are fluxes on
the real de--Rham cohomology group $H^3_{DR}$ while in the second
case the fluxes are on complex Hodge filtration of this latter:
$H^3_{DR}=H^{(3,0)}+H^{(2,1)}+H^{(1,2)}+H^{(0,3)}$. The central
charge lies in $H^{(3,0)}$, while the other charges lie
in $H^{(2,1)}$.
\par
Note that from:
\begin{equation}
\int_{S_2} T^{-} ~=~{Z}
\label{nplusone_1}
\end{equation}
it follows that ${\rm Re}\, Z, {\rm Im}\, Z$ are the electric and magnetic
charge of the graviphoton field.
The conifold points are {\it
poles} in the Yukawa couplings (Consequence of the Picard--Fuchs
equations)
and correspond to the logarithmic singularities around a
non--perturbative monopole point of the prepotential in heterotic string
considered in the previous section. Note also that the Yukawa couplings have no
monodromies since they are tensors under duality rotations.
Actually the tensors $W_{ijk}=$ $\partial_i\partial_j\partial_k {\cal F}$,
that are commonly named Yukawa couplings because of their physical
interpretation when the Calabi--Yau manifold is used to compactify the
heterotic string to an N=1 theory, in type-II N=2 compactification have the
physical interpretation of anomalous magnetic moments
of the gauginos $\lambda^i$.
\par
Using the holomorphic expression of the central charge \eqn{cc10} we may
write a general formula for the behaviour of the prepotential $F(\phi)$,
near the singular locus $Z(\phi)=0$, \cite{AFGNT}, i.e.
\begin{equation}
\label{ccsing}
F(\phi)\sim{\ii c\over\pi} Z^2(\phi) \log Z(\phi),
\end{equation}
where $c$ is a constant.
\par
The singularities of the prepotential ${\cal F}$,
that are interpreted as monopole point singularities, have their
origin in the mass singularities of these magnetic moments, that in turn
are a consequence of the Picard--Fuchs equations.
When hypermultiplets are contained in the theory, they can contribute
an anomalous magnetic moment term to $ W_{ijk} $.
\par
In the example of the quintic hypersurface \cite{candelas} in
$\IC\IP^4$, for which $h^{2,1}=1$,\cite{stromco}
\begin{equation}
\partial^3 {\cal F} \, {\stackrel{z \sim 0}{\approx}} \, {{1}\over Z}
\quad \quad \left (\mbox{in general} \,
F_{ijk} \sim {{n_i \, n_j \, n_k}\over
{n^\Lambda X_\Lambda}} \, \right )
\label{nplusone_2}
\end{equation}
which has the physical meaning of a hypermultiplet of mass $Z$
contributing to the anomalous magnetic moment of the gaugino which
is partner of the unique R--R vector field other than the graviphoton.
\par
Now, on the heterotic side, the meaning of the symplectic
section $\Omega = \left ( X^\Lambda \, , \, F_\Sigma \right )$ is
precisely the same as here, in the sense that  $\Omega$ is related
to the gauge coupling matrix ${\cal N}_{\Lambda\Sigma}$ of the heterotic
vectors by the same formulae. However the  explicit expression
of $\Omega$ looks pretty different in this case. For example on
$K_3 \times T_2$ the dependence on the vector multiplet
moduli is:
\begin{equation}
F_\Lambda \, = \, {\cal S} \, X^\Lambda \quad \quad \left(
\mbox{${\cal S}$ is the heterotic dilaton}\right)
\label{nplusone_4}
\end{equation}
at the classical level and:
\begin{equation}
F_\Lambda \, \sim \, {\cal S} \, X_\Lambda \, + \,
{\ii\beta\over\pi}n_\Lambda \, (X \cdot n) \log (X\cdot n)/X_0
\label{nplusone_5}
\end{equation}
after perturbative quantum correction,
where $\beta$ is a model-dependent constant proportional to the field
theory $\beta$-function and
$X\cdot n = 0$ is a perturbative singularity.
\section{Final Remarks}
In this paper, following a previous conjecture
\cite{CDF,CDFVP}, we have provided
a search for Calabi--Yau manifolds embedding the
R-symmetry and the quantum
monodromy of the Riemann surfaces encompassing the
non-perturbative dynamics of rigid Yang--Mills theories.
In the framework of string theory this search
finds a natural setting in dual pairing of N=2 superstring theories
\cite{FHSV,kava},
i.e.heterotic strings on $K3\times T_2$ and type-II strings on
Calabi--Yau
 threefolds, where the number of neutral massless hypermultiplets
$N_H$ of heterotic string and the vector multiplets in the
abelian phase $N_V$ match the Hodge
numbers of the dual pair according to the formula
(for type II B, for instance):
\begin{equation}
\label{r1}
N_V = h^{(2,1)} \hskip 0.3cm, \hskip 0.3cm N_H = h^{(1,1)} + 1
\end{equation}
A recent construction \cite{FHSV} of a dual pair,
based on the analysis of the soliton string worldsheet
(in the context of N=2 orbifolds of dual N=4
compactifications of type-II and heterotic strings)
and the classifications of many other candidate pairs \cite{kava}
including stringy analogue of Seiberg--Witten monopole points,
gives a further strong evidence that dynamical
Calabi--Yau manifolds,
considered in this paper purely from the point of
view of extending the quantum monodromy of the rigid theories,
are the natural candidates for describing the non-perturbative
regime of strongly coupled N=2
superstring theories, in four dimensions.
\par
We would like to finally comment on the fact that all previous analysis
are based on standard constructions of heterotic theories and Type II
theories compactified on $T_6$, $K3\times T_2$ or Calabi--Yau threefolds.
However, as recently pointed out by Chauduri and Polchinski \cite{chau1}
and Kounnas \cite{koun}, there are heterotic theories with $N=4$, $N=2$
and $N=1$ supersymmetry which do not have such geometrical interpretation
and the same is true for their Type II counterparts. Indeed, such theories
have been previously discussed in several contexts \cite{chau2,fekou}.
It would be interesting to explore weak-strong coupling duality for these
 theories and a first clue was given in \cite{chau1}. In this context one
may investigate whether in four dimensions, $N=1$ heterotic models
may be dual to $N=1$ Type II models\footnote{ After the completion of
this work, a paper has appeared in the literature \cite{vwnew} where dual
pairs of $N=1$ theories are proposed.}.
This would realize, at the string level,
the electric-magnetic duality of $N=1$ rigid theories recently explored
in the literature \cite{seibin}.
\vskip 0.1cm\noindent
\centerline{\bf Acknowledgements}
\vskip 0.1cm\noindent
We would like to thank
I. Antoniadis, P. Candelas, R. Catenacci,  R. Finkelstein,
C. Gomez, J. Harvey, A. Klemm, C. Kounnas, W. Lerche, G. Moore,
K. S. Narain, C. Reina,
R. Schimmrigk, A. Strominger, T. Taylor, C. Vafa, R. Varadarajan for
enlightening discussions.
\newpage
\appendix
\section*{Appendix. Some further consideration on the $r=1$ case}
\setcounter{equation}{0}
\addtocounter{section}{1}
Let us rewrite the potential for the $SU(2)$ dynamical Riemann
surface as follows:
\begin{equation}
0~=~{\cal W}(X,Y,Z;u)~=~ -Z^2\, + \,
{\o {1}{4}} \,
\left ( X^4 + Y^4 \right )
 \, + \, {\o{u}{2}}  X^2 Y^2
\label{diedro_1}
\end{equation}
One realizes that this potential has a
$\Gamma_{\cal W}=D_3$ symmetry group \cite{giveon,lopez}
defined by the following generators and relations
\begin{equation}
{\hat A}^2= \bfone\quad , \quad C^3=\bfone \quad, \quad
(C\hat  A)^2 =\bfone
\end{equation}
with the following action on the homogeneous coordinates
and the modulus $u$, as defined in \eqn{diedro_2}
\begin{equation}
\begin{array}{llll}
M_{\hat A} =
\left (\matrix {i & 0&0 \cr 0 & 1&0 \cr 0&0&1\cr}
\right )\ ; & \phi_{\hat A} (u)\, = \, - u \ ; & f_{\hat A}(u)\ ;\,
= \, 1& g_{\hat A}=i\\
M_C = {\frac{1}{\sqrt{2}}} \,
\left (\matrix {i & 1&0  \cr -i & 1&0  \cr 0&0&\sqrt{1+u}\cr}
\right )\ ; & \phi_C (u)\, = \,{\frac{u-3}{u+1}}\ ; & f_C(u)\, = \,
{\frac{1+u}{2}}\ ;&g_C=i\sqrt{\frac{1+u}{2}}\\
\end{array}
\label{diedro_5}
\end{equation}

Eq. (\ref{diedro_5}) is given in reference \cite{lopez}, where the
authors posed themselves the question why only the
$\ZZ_2$ cyclic group generated by $\hat A$ is actually realized as an
isometry group of the rigid special K\"ahlerian metric.
The answer is contained in the general discussion of
section~\ref{ss:monoSUr1}:
\begin{equation}
\begin{array}{ccccccc}
\ZZ_2 &=& \Gamma_{\cal W}^{rig} & \subset &
\Gamma_{\cal W} & = & D_3 \\
\end{array}
\label{diedro_7}
\end{equation}
Namely it is only $\ZZ_2$ that preserves the symplectic section
with a unit rescaling factor.
The natural question at this point is
what is the relation of this
$\ZZ_2 \subset D_3$ with the dihedral $D_4$ symmetry
expected for $r=1$. The answer is simple:
the $\ZZ_4$ action in $D_4$ becomes a $\ZZ_2$ action on the $u$
variable, $u\to \alpha^2 u$, ($\alpha^4=1$).
\vskip 0.1cm\noindent
{\bf The rigid special K\"ahler metric for $SU(2)$}
\vskip 0.1cm\noindent
As it has been shown in \cite{CDF} the Picard--Fuchs equation
associated, in the $SU(2)$ case, to the symplectic section:
\begin{equation}
\begin{array}{cccc}
\Omega_u~ =~ &\partial_u \Omega = &
\partial_u \, \left ( \matrix{Y\cr{\o{\partial {\cal F}}{\partial Y}}
\cr} \right )~=~ & \left ( \matrix{\int_A \, \omega\cr\int_B \,\omega
\cr} \right )
\end{array}
\label{diedro_8}
\end{equation}
is
\begin{equation}
\label{diedro_10}
\left ( \partial_u \bfone - A_u \right ) V = 0\ , \label{mamma2}
\end{equation}
where $V$ is defined in \eqn{V2r}, and the $2\times 2$ matrix connection
$A_u$ is given by:
\begin{equation}
\label{diedro_11}
A_u = \left(\matrix{0 & -{1\over 2}
\cr {- 1/2\over 1 - u^2}  &
{2 u\over 1 - u^2} }\right)
\end{equation}
with solutions
\begin{equation}
\label{k3t2t2}
\left\{\begin{array}{l}
{\del_u Y} \equiv f^{(1)}(u) =
F({1\over 2},{1\over 2},1;{1 + u\over 2}) +
\ii F({1\over 2},{1\over 2},1;{1 -u\over 2})\\ \null\\
\partial_u \,\frac{\del {\cal F}}{\del Y} \equiv
f^{(2)}(u) =
\ii F({1\over 2},{1\over 2},1;{1 -u\over 2}).\end{array}\right.
\end{equation}
As it is obvious, $f^{(1)}(u)$ and $f^{(2)}(u)$ just provide a
basis of two independent solutions of the linear second
order differential equation derived from the linear system
(\ref{diedro_10}). Any other pair of linear combinations of
the above functions would solve the same linear system.
The reason why precisely $f^{(1)}(u)$ and $f^{(2)}(u)$ are
respectively
identified with ${\del_u Y}$ and
$\partial_u \,\frac{\del {\cal F}}{\del Y}$  is given
by the boundary conditions imposed at infinity. When
$ u \, \rightarrow \, \infty $, the
special coordinate $Y(u)$ must approach the value it
has in the original microscopic $SU(2)$ gauge theory.
There the parameter $u$ was defined as the restriction to
the Cartan  subalgebra of the gauge invariant quadratic
polynomial
$ {\rm tr} \left ( Y^x  \sigma_x \right )^2$
so that $u={\rm const} (Y)^2$, the special coordinate
$Y(u)$ of the effective theory being $Y^3$ of the microscopic
one. Correspondingly the boundary condition at infinity
for $Y(u)$ is (with a suitable normalisation)
\begin{equation}
Y(u) \approx 2\, \sqrt{2u}+\ldots \qquad\mbox{for }u\to\infty
\label{asintuno}
\end{equation}
At the same time when $ u \, \rightarrow \, \infty $ the
non perturbative rigid special geometry
must approach its perturbative limit  defined by the following
prepotential:
\begin{equation}
{\cal F}_{pert} (Y) \, \equiv \, {{\rm i}\over {2\pi}} \,
Y^2 \, \mbox{log} \, Y^2
\label{asintpert}
\end{equation}
Combining eq.(\ref{asintuno}) and (\ref{asintpert}) we obtain
\begin{equation}
\frac{\del {\cal F}}{\del Y} \, \approx \, {{\rm i}\over {\pi}}
2\, \sqrt{2u} \, {\rm log } \, u +\ldots \qquad\mbox{for }u\to\infty
\label{asintdue}
\end{equation}
so that we can conclude:
\begin{eqnarray}
 {\del_u Y}   \approx & \sqrt{{2 \over u}}
 +\ldots \qquad & \mbox{for }u\to\infty \nonumber\\
 \partial_u \,\frac{\del {\cal F}}{\del Y}   \approx &
 {{\rm i}\over {\pi}} \,\sqrt{2\over u}
 \, {\rm log} u
 +\ldots \qquad & \mbox{for }u\to\infty\ .
\label{bordini}
\end{eqnarray}
The boundary conditions (\ref{bordini}) are just realized by
the choice of eq.s (\ref{k3t2t2}). To see this, first recall the relation
between hypergeometric functions and elliptic integrals:\footnote{We
use the notation $K(x)$ for what is usually denoted as $K(k)$ where
$x=k^2$, and similar for other elliptic integrals.}
\begin{eqnarray}
{\pi \over 2} \, F({1\over 2},{1\over 2},1;{x})&=& K(x)
\, \equiv \, \int_{0}^{\pi \over 2} \,
\left (1 - x \, \mbox{Sin}^2 \theta  \right )^{-{1\over 2}}\,  d \theta
\nonumber\\
{\pi \over 2} \, F({1\over 2},-{1\over 2},1;{x})&=& E(x)
\, \equiv \, \int_{0}^{\pi \over 2} \,
\left (1 - x \, \mbox{Sin}^2 \theta  \right )^{{1\over 2}} \, d \theta
\nonumber\\
{\pi \over 4} \, F({1\over 2},{1\over 2},2;{x})&=& B(x)=
\left ( {E(x) \over x}\,  + \, {{x-1}\over {x}} \, K(x) \right ) \ ,
\end{eqnarray}
where the square roots are positive for $0<x<1$. The functions have a
cut on the real axis from $x=1$ to $x=+\infty$. There are
the relations
\begin{eqnarray}
\int_{0}^{x} \, K(t) \, dt &=& 2\,x\, B(x)\nonumber\\
K(\ft{1}{x}) &=&\sqrt{x}\left(K(x)\pm iK(1-x)\right)\nonumber\\
 \int_0^x \ft1{\sqrt{t}} K(\ft1t) dt &=& 2\,\sqrt{x}\, E(\ft1x)\pm 2i
\ .
\end{eqnarray}
Where two signs are given, one should use them for $\mp \Im x>0$. This
indicates also on which side of the cut $(1/x)$ has to be taken when
$0<x<1$. Therefore,
the solutions of eq.s(\ref{k3t2t2}) can also be written as
\begin{equation}
\label{periodando}
\left\{\begin{array}{l}
{\del_u Y} \equiv f^{(1)}(u) =\frac{2}{\pi}\left[
K \left ({{1+u} \over 2} \right ) +
 {\rm i} \, K \left ({{1-u} \over 2} \right )\right]
= \frac{2}{\pi}
\sqrt{{2 \over {1+u}}}
\, K \left (
 {2 \over {1+u}} \right )
 \\ \null\\
\partial_u \,\frac{\del {\cal F}}{\del Y} \equiv
f^{(2)}(u) =  \frac{2}{\pi}
\ii K \left ({{1-u} \over 2} \right )\ ,\end{array}\right.
\end{equation}
where we understand here and further $+i\epsilon$ for the argument
${2 \over {1+u}} $ on the positive real axis.
By means of an integration one then obtains:
\begin{equation}
\label{integrando}
\left\{\begin{array}{l}
Y(u) = \frac{2}{\pi}\int_{u_0}^u \,\sqrt{{2 \over {1+t}}} \, K \left (
 {2 \over {1+t}} \right ) \, dt \, = \,
 {8\over {\pi}} \, \sqrt{\frac{ 1+u}{2}} \,
 E \left ( {2 \over {1+u}} \right ) \, + \, {\rm const}
 \\ \null\\
 \frac{\del {\cal F}}{\del Y}  =\frac{2}{\pi}
\ii \, \int_{0}^{u} \, K \left ({{1-t} \over 2} \right )
\, dt \, = -\frac{4i}{\pi} \,(1-u)\, B\left( \frac{1-u}{2}\right)
\, + \, {\rm const.}
\end{array}\right.
\end{equation}
Choosing zero for the integration constants, the result
(\ref{integrando}) coincides with the integral representations
originally given by Seiberg and Witten \cite{SW1}:
\begin{equation}
\label{paragonando}
\left\{\begin{array}{l}
Y(u) = 2a(u)=\frac{2\sqrt{2}}{\pi}
\int_{-1}^{1} \,\sqrt{{u-x} \over {1-x^2}} \, dx
 \\ \null\\
 \frac{\del {\cal F}}{\del Y}  =  2\,a_D(u)=2i\frac{\sqrt{2}}{\pi}
 \int_{1}^{u} \,
\sqrt{{u-x} \over {1-x^2}} \, dx
\ .\end{array}\right.
\end{equation}
The asymptotic behaviour have been given in \cite{SW1} (6.24) and
(6.26), and lead to \eqn{asintuno} and  \eqn{asintpert}.

Equipped with the above explicit solutions we can discuss
duality, monodromy, R symmetry and the explicit special
metric on the rigid special manifold.
The duality group of
electric--magnetic rotations is, in this case \cite{SW1}:
\begin{eqnarray}
\Gamma_D~& \equiv &{\bar G}_\theta \, \subset \, PSL(2,\ZZ)
\nonumber\\
{\bar G}_\theta & \equiv & \left(\matrix{1 & -1 \cr 0 &
1 \cr }\right)^{-1}  \, G_\theta \left(\matrix{1 & -1 \cr 0 &
1 \cr }\right)
\label{diedro_24}
\end{eqnarray}
namely the conjugate, via the translation matrix
$\left(\matrix{1 & -1 \cr 0 & 1}\right)$
of that subgroup $G_\theta$ of the elliptic modular group which is
generated by the two matrices $S=\left(\matrix{0 & 1 \cr -1 &
0 \cr }\right)$ and $T^{(-2)}=\left(\matrix{1 & -2 \cr 0 &
1 \cr }\right)$ (\cite{gunning}). Indeed the group ${\bar G}_\theta$
is defined by its action on the symplectic section
$\Omega_u $ which is generated by the two matrices:
\begin{eqnarray}
\label{diedro_25}
R &=& \left(\matrix{- 1 & 2 \cr -1 & 1 \cr}\right) \,  =
\, \left(\matrix{1 & -1 \cr 0 &
1 \cr }\right)^{-1}  \, S \left(\matrix{1 & -1 \cr 0 &
1 \cr }\right) \nonumber \\
T_1 &=& \left(\matrix{1& -2\cr 0& 1 \cr }\right)
\,  =
\, \left(\matrix{1 & -1 \cr 0 &
1 \cr }\right)^{-1}  \, T^{(-2)} \left(\matrix{1 & -1 \cr 0 &
1 \cr }\right)
\end{eqnarray}
where $R$ is the $R$-symmetry generator and
$T_1$ is the monodromy matrix associated with the singular point
$u=1$ of the Picard--Fuchs system (\ref{mamma2}). This
is explained as follows.
Relying on eq.(\ref{rie_3}),
we easily derive the relation between isometries
 $u_i \, \to \phi_i(u)$ of the
rigid special K\"ahler metric
and symplectic transformations \cite{noi}: there exist
$M_\phi \, \in \, Sp(2r,\IR)$ such that
\begin{equation}
\begin{array}{cccc}
\null\\
~&\Omega  \left (\phi (u)\right )\,
 &=&
e^{i\theta_\phi} \, M_\phi \, \Omega
\left (u\right )\,\\
\null\\
~&\Omega_{u_i} \left (\phi (u)\right )\,
{\o {\partial \phi^{i}}{\partial u_j}} &=&
e^{i\theta_\phi} \, M_\phi \, \Omega_{u_j}
\left (u\right )\,\\
\end{array}
\label{diedro_26}
\end{equation}
The isometry $u\to -u$, corresponding to $R$-symmetry
($Y \to \ii \, Y$ in the perturbative limit)
induces, in this theory,
the transformation
\begin{equation}
-\Omega_{u} ( - u)\, =
\mbox{i} \, \pmatrix { -1 &  2 \cr -1 &  1 \cr} \,
\pmatrix{ f^{(1)}(u) \cr f^{(2)}(u) \cr } =
\mbox{i} \, R\,  \Omega_{u}(u)
\end{equation}
while the monodromy transformation around $u=1$ gives ($r$ small)
\begin{equation}
\Omega_{u} \left (1+r\, e^{2\pi i} \right )\, =\,
 \pmatrix { 1 & -2 \cr 0 & 1} \,
\pmatrix{ f^{(1)}(1+r) \cr f^{(2)}(1+r) \cr }=
T_1\,  \Omega_{u}(1+r) \ .
\label{diedro_27quat}
\end{equation}
Having recalled the explicit form of the isometry--duality group
let us now study the structure of the rigid special metric.
To this effect let us introduce the ratio
of the two solutions to eq. (\ref{mamma2}),
\begin{equation}
\label{k3t2t5}
{ {\bar {\cal N}}}(u) = {f^{(2)}(u)\over f^{(1)}(u)}.
\end{equation}
\vskip 0.1cm\noindent
Such a ratio is identified with the matrix ${ {\bar {\cal N}}}$ appearing
in the vector field kinetic terms:
\begin{equation}
{\cal L}^{vector}_{kin}\,=\, { \o{{\rm i}}{2} } \,
[ {\bar {\cal N}}(u) \, F^{-}_{\mu\nu} \,F^{-}_{\mu\nu} \, - \,
{\cal N}({\bar u}) \, F^{+}_{\mu\nu} \,F^{+}_{\mu\nu} ]
\label{diedro_27}
\end{equation}
If we look at the {\em inverse function} $u({{\bar {\cal N}}})$,
this latter is a modular form of the group $\Gamma(2)$ that
has the following behaviour:
\begin{equation}
\begin{array}{ccc}
\forall \, \gamma \in \Gamma(2) &
u\left ( \gamma \cdot {\bar {\cal N}} \right ) =& u\left
 ( {\bar {\cal N}} \right )\\
\left ( \matrix{ -1 & 2\cr -1 & 1\cr }\right )
 \in  \Gamma / \Gamma(2) =D_3 &
 u\left ({\o{-1+\bar {\cal N}}{-1+2{\bar {\cal N}}}} \right ) =&
- \,  u\left ( {\bar {\cal N}} \right ) \\
\left ( \matrix{ 1 & 1\cr 0 & 1\cr }\right ) \, \in \,
\Gamma / \Gamma(2) =D_3 & u\left (\frac{\bar {\cal N}}
{\bar {\cal N} +1} \right )  = &
{\o{ 3-u\left ( {\bar {\cal N}} \right )   }
{ u\left ( {\bar {\cal N}} \right )  + 1 }}  \ .
\end{array}
\end{equation}
Recalling  eq.(\ref{rie_3bis})  we can now write the explicit
form of the rigid special K\"ahler metric in the variable $u$:
\begin{equation}
ds^2~=~g_{u{\bar u}} \, |du|^2 \ ;\qquad g_{u{\bar u}}=
2 \, {\rm Im}\, {\bar {\cal N}}(u) \, |f^{(1)}(u)|^2\\
\label{metricozza}
\end{equation}
\par
Calculating the Levi--Civita connection and Riemann tensor of
this metric we obtain:
\begin{equation}
\begin{array}{ccccc}
\Gamma^{u}_{u u}&=&-\, g^{u{\bar u}} \,
\partial_u \, g_{u{\bar u}} &=&-{\o{1}{2 {\rm i}}} \,
{\o { \partial {\bar {\cal N}}/\partial u}{ {\rm Im}\,
 {\bar {\cal N}}(u) } }\, - \,
\partial_u \mbox{log} f^{(1)}(u) \\
\null\\
R^{u}_{u{\bar u }u}&=& \,
\partial_{\bar u} \,\Gamma^{u}_{u u} &=&{\o{1}{4 }} \,
{\o {1 }{ ({\rm Im}\, {\bar {\cal N}}(u)})^2 } \,
|\partial {\bar {\cal N}}/\partial u |^2\\
\null\\
R_{{\bar u}u{\bar u} u}&=& \,
g_{u{\bar u}} R^{u}_{u{\bar u}u }&=&{\o{1}{2 }} \,
{\o {1 }{ {\rm Im} \,{\bar {\cal N}}(u)} } \,
|\partial {\bar {\cal N}}/\partial u |^2 \, |f^{(1)}(u)|^2\\
\end{array}
\label{curvelievi}
\end{equation}
so that we can verify that the above metric is
indeed {\it rigid special
K\"ahlerian}, namely that it satisfies the constraint:
\begin{equation}
R_{{\bar u}u{\bar u} u}\, - \, C_{uuu} \,
{\bar C}_{{\bar u}{\bar u}{\bar u}} \, g^{u{\bar u}}~=~0
\label{costretto_1}
\end{equation}
by calculating the Yukawa coupling or anomalous magnetic moment
tensor:
\begin{equation}
C_{uuu}~=~ \partial_u {\bar {\cal N}} \,
\left ( f^{(1)}(u) \right )^2
\label{ctensore}
\end{equation}
As one can notice from its explicit form (\ref{metricozza}),
the K\"ahler metric of the rigid N=2 gauge theory of rank
$r=1$ is not the Poincar\'e metric in the variable
${\bar {\cal N}}$, as one might naively expect from the fact that
${\bar {\cal N}}=\tau$ is the standard modulus of a torus and that
$G_{\theta}\, \subset \, PSL(2,\ZZ)$ linear fractional
transformations are isometries. Indeed
(\ref{metricozza})
is to be contrasted with the expression for the
Poincar\'e metric:
\begin{equation}
ds^2~=~g^{Poin}_{{\cal N}{\bar{\cal N}}}\,
 |d {\bar {\cal N}} \, |^2
~=~ \, {\o{1}{4}} \,
{\o{1}{({\rm Im}\, {\bar {\cal N}})^2}} \, |d {\bar {\cal N}} \, |^2
\ . \label{metricapoincare}
\end{equation}
{}From eq.(\ref{curvelievi}) however it is amusing to note that the Ricci
form of the rigid metric is precisely the Poincar\'e metric.
\begin{equation}
R^{Ricci}_{~~~~{{\cal N}}{\bar { {\cal N}}}}~=~
g^{Poin}_{~~~~{ {\cal N}}{ {\bar {\cal N}}}}
\label{masterequation}
\end{equation}
This is a consequence of the general equation \eqn{boh} in the case of
 one modulus where the period matrix $\cal N$ can be used as a parameter.
\vskip 0.1cm\noindent
{\bf The rigid special coordinates}
\vskip 0.1cm\noindent
In the special coordinate basis the anomalous magnetic moment tensor
is given by:
\begin{equation}
C_{YYY}~=~C_{uuu} \, \left ( {\o{\partial u}{\partial Y}} \right )^3
~=~-{{2\ii} \over \pi}
{\o{1}{1-u^2}} \, \left ( {\o{\partial u}{\partial Y}} \right )^3
\label{cyyy}
\end{equation}
The second of equations (\ref{cyyy}) follows from the comparison
between equation (\ref{ctensore}) and the Picard--Fuchs
equation (\ref{diedro_10}) satisfied by the periods that yields:
\begin{equation}
C_{uuu}~=~-{{2\ii} \over \pi} \, {\o{1}{1-u^2}}
\end{equation}
In the large $u$ limit the asymptotic behaviour of the special
coordinate is given by eq. (\ref{asintuno}):
\begin{equation}
C_{YYY}(u)~=~{\o{\partial^3 {\cal F}}{\partial Y^3}}(u)~
 \approx~{\o{\ii}{\sqrt{2}\pi}} \,  u^{-1/2}+\ldots
\qquad\mbox{for }u\to\infty
\label{asintoto_100}
\end{equation}
and by triple integration one verifies consistency with
the asymptotic behaviour
of the prepotential ${\cal F}(Y)$ (\ref{asintpert}):
\begin{equation}
{\cal F}(Y)~\approx~
 \frac{i}{2\pi} \, Y^2  \log {Y^2}  \, +
\ldots \qquad\mbox{for }Y\to\infty
\label{asintoto_2}
\end{equation}
Formula (\ref{asintoto_2}) contains the leading classical
form of ${\cal F}(Y)$ plus the first perturbative correction
calculated with standard techniques of quantum field--theory.
Eq.(\ref{asintoto_2}) was the starting point of the analysis
of Seiberg and Witten who from the perturbative
singularity structure inferred the monodromy group and then
conjectured the dynamical Riemann surface. The same procedure
has been followed to conjecture the dynamical Riemann surfaces
of the higher rank gauge theories.
The nonperturbative solution is given by
\begin{equation}
{\cal F}(Y)~=~  \frac{i}{2\pi} Y^2  \log \frac{Y^2}{\Lambda^2}
+Y^2\sum_{n=1}^{\infty}C_n  \left(
\frac{\Lambda^2}{Y^2}\right) ^{2n}
\label{generalformofeffe}
\end{equation}
The infinite series in (\ref{generalformofeffe}) corresponds
to the sum over instanton corrections of all instanton--number.
\par
The important thing to note is that the special coordinates $Y^\alpha(u)$
of rigid special geometry approach for large values of $u$ the
Calabi--Visentini coordinates of the manifold $O(2,n)$ $/O(2)\times$
$O(n)$ discussed in  section~\ref{ss:SGCY}.
As stressed there, the $Y^\alpha$ are
not special coordinates for local special geometry.

\end{document}